\documentclass{vldb-noconfinfo}

\usepackage{amsmath,amssymb}
\usepackage[linesnumbered,ruled,vlined]{algorithm2e}
\SetKwComment{tcp}{$\triangleright$\ }{}
\usepackage{threeparttable}
\usepackage{paralist}
\usepackage[dvipsnames]{xcolor}
\usepackage{xspace}
\usepackage{balance}
\usepackage{ntheorem}
\usepackage{paralist}
\usepackage{hyperref}
\usepackage[nocompress]{cite}  %
\newcommand{\tsf}{\textsf}

\newcommand{\myparagraph}[1]{\noindent\textbf{#1}.\,}

\newcommand{\revise}[1]{#1} %

\newcommand{\confversion}[1]{}
\newcommand{\fullversion}[1]{#1}
\newtheorem{theo}{Theorem}
\newtheorem{defi}{Definition}
\newtheorem{prop}{Proposition}
\newtheorem{lemm}{Lemma}
\newtheorem{exam}{Example}

\newcommand{\SigmaEps}{\Sigma^+}  %
\newcommand{\subseq}{\subseteq}
\newcommand{\substr}{\sqsubseteq}

\newcommand{\trie}{\mathcal{T}}
\newcommand{\database}{\mathcal{T}}

\newcommand{\wed}{\textsf{wed}\xspace}
\newcommand{\WED}{WED\xspace}
\newcommand{\WEDs}{WEDs\xspace}
\newcommand{\sub}{\textsf{sub}\xspace}
\newcommand{\ins}{\textsf{ins}\xspace}
\newcommand{\del}{\textsf{del}\xspace}

\newcommand{\edr}{\textsf{EDR}\xspace}
\newcommand{\erp}{\textsf{ERP}\xspace}
\newcommand{\edwp}{\textsf{EDwP}\xspace}
\newcommand{\lev}{\textsf{Lev}\xspace}

\newcommand{\surs}{\textsf{SURS}\xspace}
\newcommand{\sursfull}{shortest unshared road edges\xspace}
\newcommand{\neterp}{\textsf{NetERP}\xspace}
\newcommand{\netedr}{\textsf{NetEDR}\xspace}

\newcommand{\dtw}{\textsf{DTW}\xspace}
\newcommand{\lcss}{\textsf{LCSS}\xspace}
\newcommand{\lors}{\textsf{LORS}\xspace}
\newcommand{\lcrs}{\textsf{LCRS}\xspace}

\newcommand{\dist}{d}

\newcommand{\topk}{top-$k$\xspace}

\newcommand{\filtername}{subsequence filtering\xspace}
\newcommand{\Filtername}{Subsequence Filtering\xspace}
\newcommand{\GSF}{OSF}  %

\newcommand{\beijing}{\tsf{Beijing}\xspace}
\newcommand{\porto}{\tsf{Porto}\xspace}
\newcommand{\singapore}{\tsf{Singapore}\xspace}
\newcommand{\sanfran}{\tsf{SanFran}\xspace}

\newcommand{\dita}{\tsf{DITA}\xspace}

\newcommand{\size}[1]{|#1|}
\newcommand{\set}[1]{\{#1\}}

\usepackage[all=normal, floats, bibnotes, wordspacing, charwidths, indent, lists]{savetrees}

\makeatletter%
\skip\footins 8.1pt plus 4pt minus 2pt %
\vldbTitle{Fast Subtrajectory Similarity Search in Road Networks under Weighted Edit Distance Constraints}
\vldbAuthors{Satoshi Koide, Chuan Xiao, Yoshiharu Ishikawa}
\vldbDOI{https://doi.org/10.14778/3407790.3407818}
\vldbVolume{13}
\vldbNumber{11}
\vldbYear{2020} 
\begin{document}

\pagenumbering{arabic}
\confversion{\pagestyle{empty}}
\fullversion{\pagestyle{plain}}

\title{Fast Subtrajectory Similarity Search in Road Networks under Weighted Edit Distance Constraints} %

\numberofauthors{3} %

\author{
\alignauthor
Satoshi Koide\\
       \affaddr{Toyota Central R\&D Labs.}\\
       \email{koide@mosk.tytlabs.co.jp}
\alignauthor
Chuan Xiao\\
       \affaddr{Osaka University and Nagoya University}\\
       \email{chuanx@nagoya-u.jp}
\alignauthor
Yoshiharu Ishikawa\\
       \affaddr{Nagoya University}\\
       \email{ishikawa@i.nagoya-u.ac.jp}
}

\maketitle

\begin{abstract}
In this paper, we address a similarity search problem for spatial trajectories in road networks.
In particular, we focus on the \emph{subtrajectory similarity search} problem, which involves 
finding in a database the subtrajectories similar to a query trajectory. A key feature of our 
approach is that we do not focus on a specific similarity function; instead, we consider 
\emph{weighted edit distance} (WED), a class of similarity functions which allows user-defined 
cost functions and hence includes several important similarity functions such as \edr and \erp. 
We model trajectories as strings, and propose a generic solution which is able to deal with any 
similarity function belonging to the class of WED. 
By employing the filter-and-verify strategy, we introduce \emph{\filtername} to efficiently 
prunes trajectories and find candidates. In order to choose a proper subsequence to optimize the 
candidate number, we model the choice as a discrete optimization problem (NP-hard) 
and compute it using a 2-approximation algorithm. To verify candidates, we design \emph{bidirectional 
tries}, with which the verification starts from promising positions and leverage the shared segments 
of trajectories and the sparsity of road networks for speed-up. Experiments are conducted on large 
datasets to demonstrate the effectiveness of WED and the efficiency of our method for various 
similarity functions under WED.  \end{abstract}

\section{Introduction}
\label{sec:intro}
Vehicular transportation is facing a crucial turning point as data-driven information technology advances. Data-driven approaches, such as intelligent routing and ride-sharing, are expected to resolve important social issues, such as environmental problem and traffic congestion; they are therefore actively studied in many fields, including database research. 
Accordingly, fundamental operations, such as indexing and retrieval, on huge vehicular trajectory data are becoming increasingly important~\cite{SIGIR18-Torch,SIGMOD18-DITA,ICDE18-DISON,TSAS18-SNT,SIGSPATIAL14-SPQ,SPNET,CiNCT}.

\begin{figure}[t]
    \centering
    \includegraphics[scale=0.8,trim=0 1.em 0 0]{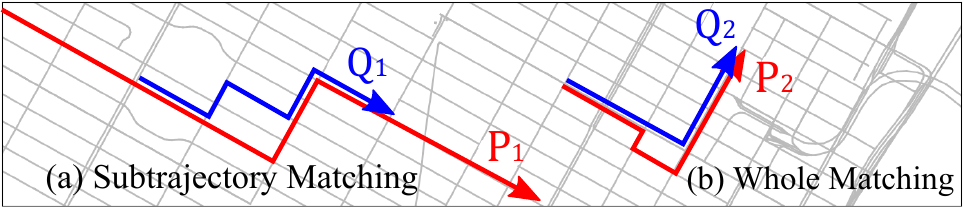}
    \caption{Subtrajectory matching/Whole matching.} 
    \label{fig:subtraj_search}
\end{figure}

This paper addresses the similarity search problem over trajectories in road networks, which is a classical but still active area of spatial database research \cite{SIGIR18-Torch,SIGMOD18-DITA,ICDE18-DISON,Xia2011,Won2009}. Unlike most existing studies that target \emph{whole matching} between data and query trajectories (i.e., 
the entire trajectories are similar), we tackle the \emph{subtrajectory} similarity search problem, which finds in a vehicular trajectory database the subtrajectories 
similar to a query (Figure~\ref{fig:subtraj_search}). 

\noindent\textbf{Why subtrajectory similarity search?}\,  One motivating application is travel time estimation along a given path. Recent on-the-fly approaches~\cite{MDM17-Waury,EDBT19-Waury} estimate the travel time distribution by retrieving historical trajectories that contain the query path as a subtrajectory immediately after the query arrived. 
\revise{Other applications include alternative route suggestion that finds the variations of a query path in the database as alternative routes} and 
path popularity estimation~\cite{TSAS18-SNT,SIGMOD13-TPMFP,ICDE10-MPR} that counts the frequency of appearance of a given path in the database as a 
subtrajectory.

Exact path queries have been studied for subtrajectory search~\cite{SIGSPATIAL14-SPQ,TSAS18-SNT}; 
however, exact path queries only find trajectories containing a subtrajectory that exactly 
matches a query trajectory. %
Hence similarity queries are adopted to retrieve more semantically relevant results for 
various applications~\cite{DBLP:conf/icde/YiJF98,Keogh2005,ERP,EDR,FeiFei2017,ShuoShang2017,SIGIR18-Torch,SIGMOD18-DITA,ICDE18-DISON,VLDBJ-survey}. 
For example, similarity queries can handle the errors caused by sampling strategies, spatial 
transformations, or natural noises~\cite{EDR,ERP,SIGIR18-Torch,ICDE18-DISON,VLDBJ-survey}, 
which are common in real data. In addition, travel time estimation suffers from \emph{data sparsity} 
(i.e., there are few historical trajectories that exactly travel a query path in the specified time slot) 
even in urban areas. Subtrajectory similarity queries have been used to address this issue; e.g., by path 
similarity~\cite{DBLP:conf/sdm/IdeK09} or road segment similarity (types, number of lanes,  etc.)~\cite{DBLP:conf/kdd/WangZX14}, 
based on the observation that paths or road segments with similar contexts may have similar travel time.

\myparagraph{Trajectory similarity functions} To measure the similarity between trajectories, a similarity 
function must be selected. Albeit many trajectory similarity functions have been proposed \cite{DBLP:conf/icde/YiJF98,Shim2003,EDR,ERP,FeiFei2017,ShuoShang2017,VLDB14-Personalized,SIGIR18-Torch,Won2009,ICDE18-DISON,Xia2011,Evans2013,Hwang2006,Tiakas2009}, as demonstrated experimentally~\cite{SIGIR18-Torch,ICDE18-DISON,EDR,ERP}), every similarity function has its own advantages and disadvantages.
In other words, there is no ``best'' trajectory similarity function, and the choice depends on the application scenario. 
Accordingly, for trajectory similarity search, general methodologies that do not depend on a specific similarity function are preferable.

In this paper, we consider \emph{weighted edit distance} (WED), a class of similarity functions that includes 
several ones commonly used in trajectory analysis, such as edit distance on real sequences (\edr) \cite{EDR} and 
edit distance with real penalty (\erp) \cite{ERP}, \revise{which have been shown capable of handling different 
sampling strategies, spatial transformations, and/or noises, and better than other functions such as dynamic time 
warping (\dtw) and longest common subsequence (\lcss)~\cite{VLDBJ-survey}.} 
WED is \emph{flexible} in the sense that it allows user-defined 
cost functions. As such, it not only covers \edr and \erp but also their extensions (e.g., the adaptations using road 
network distance instead of binary or Euclidean distance). 
WED also captures the semantics of the aforementioned similarities 
for travel time estimation~\cite{DBLP:conf/sdm/IdeK09,DBLP:conf/kdd/WangZX14}, and is able to 
measure the road segments that differ between two trajectories, thereby (reversely) expressing 
semantics similar to longest overlapping road segments (\lors)~\cite{SIGIR18-Torch} and longest 
common road segments (\lcrs)~\cite{ICDE18-DISON}.

\myparagraph{Challenges} 
\revise{For real applications like trajectory analysis~\cite{SIGMOD18-DITA,ICDE18-DISON}, it is 
desirable to return query results in seconds or even less. Trajectories in road networks can be 
modeled as strings, and the problem is 
converted to substring search. A naive approach is scanning the database using the 
Smith-Waterman algorithm~\cite{SW}; however, this is inefficient for large datasets as it does not 
employ indexing. For fast query processing, most string similarity algorithms resort to 
the \emph{filter-and-verify} paradigm, which finds a set of candidates and then verify them. The most 
widely used is $q$-gram filtering~\cite{DBLP:conf/vldb/GravanoIJKMS01,Ed-join,DBLP:journals/tods/Qin0XLLW13,DBLP:conf/sigmod/DengLF14,DBLP:journals/tkde/WeiYL18} (also for trajectories~\cite{EDR}). However, it targets Levenshtein 
distance (unit cost) and does not deliver efficient performance for WED, because the lower bound 
of common $q$-gram number become very loose (even $\leq 0$ and thus useless) when substitution 
cost is arbitrarily small (e.g., \erp). Partition-based~\cite{DBLP:conf/sigmod/WangXLZ09,DBLP:journals/tods/LiDF13} 
and trie-based~\cite{DBLP:journals/vldb/FengWL12} methods are not applicable either, because it is 
hard to derive the partition size for WED, and they are designed for whole matching. Another key 
property resides in the sparsity of road networks (i.e., the alphabet is very large but 
spatially restricted), which has not been exploited in these solutions.}
Another line of work is trajectory similarity methods~\cite{SIGIR18-Torch,SIGMOD18-DITA,ICDE18-DISON,ERP}. 
However, these techniques are designed for whole matching and become inefficient or inapplicable on 
subtrajectories. Many of them rely on their own similarity functions (e.g., the 
\tsf{ERP-index}~\cite{ERP} exploits the triangle inequality) or need adaptations to switch between 
functions (e.g., \dita~\cite{SIGMOD18-DITA} requires a pivoting strategy depending on the similarity 
function). \revise{Moreover, all the aforementioned methods bear no performance guarantee on 
candidate size.} 
Seeing these challenges, we aim to design a \emph{unified} yet \emph{efficient} algorithm 
applicable to a wide range of similarity functions under the class of WED.%

\myparagraph{Contributions}
Our contributions are as follows: %
\begin{itemize} 
  \item We propose the first indexing and retrieval method for trajectories in road networks that supports 
  subtrajectry search on WED. The query processing algorithm does not depend on a specific similarity 
  function; it supports any user-specified edit distance with a \textbf{unified} and exact algorithm such 
  that there is no need to adapt the algorithm to switch between similarity functions. 
  To tackle the efficiency issue, we model trajectories as strings and follow the filter-and-verify paradigm. 
  \item To generate candidates, we propose the \textbf{\filtername} (\S~\ref{sec:filter}), such that any 
  trajectory in the result set must share at least one element, or a neighbor (in terms of the cost functions) 
  of the element, with a chosen subsequence of the query. In order to choose a subsequence to optimize the 
  candidate size, we model this as a discrete optimization problem and show its \emph{NP-hardness} as well as 
  a polynomial-time \emph{2-approximation} algorithm. We also give the condition under which this algorithm 
  finds the \emph{optimal} subsequence. 
  Indexing and search algorithms (\S~\ref{sec:index}) are devised based on this filtering strategy. 
  \item To verify candidates (\S~\ref{sec:verify}), we start with the positions at which candidates are found 
  and develop pruning techniques for a \textbf{local verification} algorithm, so that only the promising part 
  of each candidate is computed for WED. We share computation for the common subtrajectories of candidates, 
  and design \textbf{bidirectional tries} to cache such computation by exploiting the sparsity of road 
  networks which leads to \emph{low cache miss rate}. 
  \item We conducted extensive experiments on large real datasets (\S~\ref{sec:experiments}). The results 
  demonstrate the effectiveness of WED and the efficiency of our method on various similarity functions under 
  the class of WED, as well as the effectiveness of the components in the proposed solution. 
\end{itemize}

\section{Preliminaries}
\label{sec:prelim}
\subsection{Framework and Data Model}
\label{sec:sec:datamodel}

We assume trajectories are constrained in a road network.
A query essentially consists of three components: (i)~a trajectory, 
(ii)~a distance function, and
(iii)~a distance threshold.
On receiving a query, we find the data trajectories that 
approximately (satisfying the 
distance constraint) contains the query as a subtrajectory. 
In this paper, we assume datasets and indexes are stored in main memory. 
\fullversion{
We also focus on the single-core case and leave the distributed 
case to future work due to the challenge of partitioning for subtrajectory 
search (the partitioning methods based on first and last points of 
trajectories~\cite{SIGMOD18-DITA,ICDE18-DISON} were designed for whole 
matching and do not apply here).
}
The road network is modeled as a directed graph $G=(V,E)$.
Each vertex $v\in V$ is associated with its coordinate in $\mathbb{R}^2$.
Each edge $e\in E$ is 
associated with a weight (e.g., travel time or distance in a road network) 
denoted by $w(e)$.

A trajectory is modeled as a path on $G$, i.e., a consecutive sequence $v_1v_2\cdots v_n$ of vertices.
We refer to this trajectory representation as \emph{vertex representation}.
Equivalently, the path can be represented by the corresponding \emph{edge representation} $e_1e_2\cdots e_{n-1}$, where $e_i=(v_i,v_{i+1})\in E$. 
These representations can be converted from raw trajectory (a sequence of spatial coordinates) 
through map matching (we employ the HMM map matching for this purpose \cite{HM4}). 
The techniques proposed in this paper can support both representations. 

Timestamps are associated with each trajectory.
Following the trajectory models in existing studies (e.g., \cite{SIGSPATIAL14-SPQ,TSAS18-SNT}), we assume that timestamps are recorded at each vertex.
In summary, we employ the following definition of trajectories:

\begin{defi}[Trajectory]
A trajectory is a tuple $(P,T)$, where $P$ is a path on $G$, and $T$ is a sequence of timestamps associated with each vertex in $P$.
\end{defi}

We first focus on dealing with paths and then extend our techniques to the case with time constraints. Thus, we also denote a trajectory by its path $P$. 

\fullversion{
\begin{table}[t]
    \centering
    \small
    \caption{Frequently used notation.}%
    \begin{tabular}{c|l}\hline
    $G=(V,E)$ & Road network (directed graph)\\
    $\Sigma, \SigmaEps, \Sigma^*$     & Alphabet ($V$ or $E$), $\Sigma\cup\{\varepsilon\}$, possible strings on $\Sigma$ \\
    $q$ & A symbol \\
    $P, Q\in\Sigma^*$ & A data trajectory, a query trajectory \\
    $\database$ & A dataset of trajectories \\
    $P_i$, $P_{i:j}$ & $i$-th element, a subtrajectory (substring) from $i$ to $j$ \\
    $P'\subseq P$ & $P'$ is a subsequence of $P$ \\
    $P'\substr P$ & $P'$ is a subtrajectory of $P$ \\
    $\d(a,b), w(e)$ & Distance between $a$ and $b$, weight of $e\in E$ \\%
    $\tau$ & A distance threshold \\
    $B(q), B(Q)$ & Substitution neighbors of $q\in\Sigma$ and $Q\in\Sigma^*$ (\S~\ref{sec:sec:gsf}) \\ %
    $[\![n]\!]$ & A set of integers $[1,2,\cdots,n]$\\
    \hline
    \end{tabular}
    \label{table:notation}
\end{table}
} 
\myparagraph{Notation}
A path can be regarded as a string. 
We denote an alphabet set by $\Sigma$.
For vertex representation, $\Sigma=V$.
For edge representation, $\Sigma=E$.
The set of all possible strings on $\Sigma$ is denoted by $\Sigma^*$.
An empty symbol is denoted by $\varepsilon$, and $\SigmaEps:=\Sigma\cup\{\varepsilon\}$.
Given a trajectory $P$, its $i$-th element is $P_i$, and a subtrajectory (substring) of $A$ from 
$i$ to $j$ is $P_{i:j}$ (if $i>j$, $P_{i:j}$ represents an \emph{empty string}). 
$\size{P}$ denotes the length of the trajectory (string). 
We say $P'\substr P$ if $P'$ is a \emph{subtrajectory} of $P$.
Similarly, $P'\subseq P$ means that $P'$ is a \emph{subsequence} of $P$; 
i.e., there exist $i_1, i_2, \cdots, i_k$ such that 
$i_1<i_2<\cdots<i_k$ and $P'=P_{i_1}P_{i_2}\cdots P_{i_k}$.
\revise{$\set{1,2,\cdots,n}$ is denoted by $[\![n]\!]$.}
\fullversion{The frequently used notation in this paper is summarized in Table~\ref{table:notation}.}

\subsection{Weighted Edit Distance}
\label{sec:prelim:ged}

\subsubsection{Concept}
The Levenshtein distance, the most fundamental form of edit distance (on strings), 
counts the minimum number of \emph{edit operations} needed to convert a 
string $P$ into another string $Q$. The edit operations usually consists 
of \emph{insertion}, \emph{deletion}, and \emph{substitution} of a symbol.

We consider a general class of edit distances where the 
costs of edit operations can take any values. 
Given two symbols $a, b\in\Sigma$, we denote the insertion, deletion and 
substitution costs by $\ins(a)$, $\del(b)$, and $\sub(a,b)$, respectively.
The weighted edit distance (\WED) between two trajectories $P=P_{1:m}$ and 
$Q=Q_{1:n}$, denoted by $\wed(P,Q)$, is defined recursively:
\begin{align*}
&\hspace{-0.5em}
\wed(\varepsilon,Q_{1:n})=\sum_{j=1}^n\ins(Q_j),\quad
\wed(P_{1:m},\varepsilon)=\sum_{i=1}^m\del(P_i),\\
&\hspace{-0.5em}
\wed(P_{1:m},Q_{1:n})=\min
\begin{cases}
\wed(P_{1:m-1},Q_{1:n-1})+\sub(P_m,Q_n),\\
\wed(P_{1:m-1},Q_{1:n})+\del(P_m),\\
\wed(P_{1:m},Q_{1:n-1})+\ins(Q_n).
\end{cases}
\end{align*}
We can compute $\wed(P_{1:m},Q_{1:n})$ by dynamic programming in $O(mn)$ time.
Furthermore, to keep notation simple, we define $\sub(a,\varepsilon):=\del(a)$ and $\sub(\varepsilon,b):=\ins(b)$.
\smallskip

\myparagraph{Assumptions}
To obtain a meaningful similarity function, we make some assumptions on the edit operation costs.
First, to make $\wed(P,Q)$ nonnegative, we assume $\sub(a,b)\ge0$ for $a,b\in\SigmaEps$.
To make $\wed(P,Q)$ symmetric, %
we assume $\sub(a,b)=\sub(b,a)$ (and this implies $\ins(a)=\del(a)$).
Finally, to make $\wed(P,P)=0$ hold, we assume $\sub(a,a)=0$.
Note that we do not enforce the triangle inequality $\wed(P,Q)\le \wed(P,R)+\wed(R,Q)$.
Also, we do not enforce $\wed(P,Q)=0 \Rightarrow P=Q$, i.e., $\wed(P,Q)=0$ does not mean $P=Q$.

\begin{prop}\label{prop:ged}
With the assumptions above, we have:\\ %
    \emph{(i)} $\wed(P,Q)\ge0$, (nonnegativity);
    \emph{(ii)} $\wed(P,P)=0$, (pseudo-positive definite);
    \emph{(iii)} $\wed(P,Q)=\wed(Q,P)$. (symmetry)
\end{prop}

Proposition~\ref{prop:ged} implies that \WED is not metric in general.
Next, we show that \WED contains some existing similarity 
functions as special cases. %

\subsubsection{Known Instances of WED}
\myparagraph{Levenshtein Distance}
The well-known Levenshtein distance (\lev) is obtained by setting
\begin{align}
    \sub(a,b)=\begin{cases}
    0,&(a=b)\\
    1,&(a\ne b)
    \end{cases},\;
    \ins(a)=1,\; \del(b)=1.
\end{align}
This can be used for both the vertex and edge representations.
\smallskip \\
\myparagraph{Edit Distance on Real Sequence (\edr)}
\edr \cite{EDR} is defined on real-valued sequences. For the data trajectory $P$, we use vertex representation. %
A query $Q$ is not necessarily restricted on road networks.
We can cover \edr by setting
\begin{align}
    \hspace{-1em}\sub(a,b)=\begin{cases}
    0,&(\dist(a,b)\le \varepsilon)\\
    1,&(otherwise)
    \end{cases},\;
    \ins(a)=1,\; \del(b)=1,
    \label{eq:def-edr}
\end{align}
where $\varepsilon>0$ is a predefined matching threshold. We employ Euclidean distance for $\dist(a,b)$.
\edr is not a metric, i.e., the triangle inequality does not hold.
\smallskip \\
\myparagraph{Edit Distance with Real Penalty (\erp)}
\erp \cite{ERP} is also defined on real-valued sequences, obtained by setting
\begin{align}
    \sub(a,b)=\dist(a,b),\;
    \ins(a)=\dist(a,g),\; \del(b)=\dist(b,g),
    \label{eq:def-erp}
\end{align}
where $g\in\mathbb{R}^2$ is a predefined reference point (e.g., the barycenter of the vertices in $V$).
\erp is a metric. 

\subsubsection{Network-aware Similarity Functions}
\label{sec:prelim:ged:other}
WED is more flexible than the aforementioned distance functions in the 
sense that users can define their own costs tailored to the application; e.g., for trajectory 
analysis in a road network, users may use a distance defined on the road network instead of 
Euclidean distance, or count the road segments that differ in two trajectories. Next we 
show some examples. 
\smallskip

\myparagraph{\neterp~and \netedr}
A popular trajectory similarity definition employs shortest path distance
between two vertices $a$ and $b$~\cite{ShuoShang2017,VLDB14-Personalized,Evans2013,Hwang2006,Tiakas2009}.
By replacing Euclidean distance in \edr (Eq.(\ref{eq:def-edr})) and \erp (Eq.(\ref{eq:def-erp})) with 
shortest path distance, %
we obtain new similarity functions, referred to as \netedr~and \neterp, respectively.
For directed graphs, shortest path distance %
is not symmetric, %
which violates the assumption above.
One way to fix this is to make the road network undirected.

In \erp, we need a reference point; in \neterp, we use a constant insertion/deletion cost instead, 
$G^{\tsf{(del)}}_\neterp>0$, defined by users. 
This makes \neterp~non-metric, but this does not affect our method since it does not use the triangle inequality.
\smallskip

\myparagraph{Shortest Unshared Road Segments (\surs)}
Another idea behind several existing similarity functions is to evaluate the total edge weights (e.g., distance or travel time) that are shared (or unshared) between two trajectories~\cite{Xia2011,Won2009,SIGIR18-Torch,ICDE18-DISON}. 
To express such semantics using \WED, we define \emph{\sursfull}~(\surs) for 
trajectories in edge-representation: %
\begin{align}
    \sub(a,b)=w(a)+w(b),\;
    \ins(a)=w(a),\; \del(b)=w(b),
\end{align}
where $w(a)$ is a given travel cost for a road edge $a\in E$.
Because $\sub(a,b)=\ins(a)+\del(b)$, substitution is equivalent to a combination of insertion and deletion; therefore, \surs~essentially counts the total travel costs of edges not shared between two trajectories, considering the order of sequence elements.
Note that the functions in \cite{Xia2011,Won2009,SIGIR18-Torch,ICDE18-DISON} measure similarity, 
while \surs measures distance. 
\begin{exam}
Given two paths $P=\texttt{befg}\in \Sigma^*$ and $Q=\texttt{abcdg}\in \Sigma^*$
the optimal alignment that yields the \surs~is:
\begin{align*}
 \texttt{- b - - e f g}&\quad(\,=P)\\%
 \texttt{a b c d - - g}&\quad(\,=Q).
\end{align*}
$\surs(P,Q)$ is the total cost of the edges aligned to the gap symbol, i.e., $w(\texttt{a})+w(\texttt{c})+w(\texttt{d})+w(\texttt{e})+w(\texttt{f})$.%
\end{exam}

So far we have discussed WED instances. Furthermore, given a supervised machine learning task, we may 
optimize the edit operation costs of WED using a technique in \cite{DBLP:conf/nips/KoideKK18}.

\subsubsection{Other Similarity Functions}
There are also other similarity functions not belonging to \WED. For example, in \dtw, one element 
of a trajectory can be aligned to multiple elements in the other; this is not allowed in \WED. 
\lcss, \lors, and \lcrs are not \WED either, because they are measure common subsequence rather than 
distance. 

\subsection{Problem Setting}
\label{sec:prelim:problem}
To give a formal definition of our subtrajetory search problem, we first define the term subtrajectory matching.

\begin{defi}[Subtrajectory Matching]
\label{defi:match}
Given a query $Q\in\Sigma^*$ and a trajectory $P\in\Sigma^*$, we say a subtrajectory $P_{i:j}$ of $P$ matches 
$Q$ (and vice versa) iff $\wed(P_{i:j},Q)<\tau$, where $\tau$ is a threshold.
\end{defi}

\begin{exam}
Consider a trajectory $P=\texttt{ABCDE}$.
As $|P|=5$, there are $|P|(|P|+1)/2=15$ subtrajectories.
Consider a query $Q=\texttt{BFD}$ under \lev with $\tau=2$.
Then, $P_{2:4}=\texttt{BCD}$ satisfies $\wed(P_{2:4},Q)=1<\tau$ and thus matches $Q$ 
(note that we use ``$<$'' not ``$\leq$'' in the problem definition).
\end{exam}

We denote a set of $N$ data trajectories by $\database=\{(P^{(id)}, T^{(id)}\}_{id=1}^N$.
Our problem is defined as follows.

\begin{defi}[Subtrajectory Similarity Search] 
\label{defi:sub-sim-search}
Given $(Q, \database, \wed, \tau)$, find in $\database$ the subtrajectories that match $Q$, i.e., 
\begin{align}
    \hspace{-.8em}\tsf{SubtrajSimSearch}(Q, \database, \wed, \tau):=\{(id, s, t) \mid P_{s:t}^{(id)} \; \text{matches} \; Q \}.
    \nonumber %
\end{align}
\end{defi}
\revise{To avoid the case that $Q$ is similar to an empty trajectory (i.e., $\wed(Q, \varepsilon) < \tau$), 
we assume that $\sum_{q \in Q} \ins(q) \geq \tau$ for a meaningful problem definition.}

\myparagraph{Temporal Constraints}
Some applications require consideration of temporal condition.
For the on-the-fly travel time estimation mentioned in \S~\ref{sec:intro}, searching trajectories that traveled during a given time interval, say $I$, is important (e.g., rush hour).
This condition can be written as $[T_i,T_j]\subseteq I$, or $[T_i,T_j]\cap I\ne\emptyset$, where $i$ and $j$ are the matched positions in Definition~\ref{defi:match}.
Another application may require constraints on average speed (i.e., $D(P_{i:j})/(T_j-T_i)$ where $D$ is the distance of $P_{i:j}$) 
or travel time for the query or some of the road segments. 

A simple and general approach is checking temporal constraints \emph{after} solving the subtrajectory similarity search. 
This allows us to treat \emph{any kind of} temporal constraints.
For some cases, however, speed-up can be achieved by considering temporal conditions during the filtering step, as discussed in \S~\ref{sec:index:extension}. 
\section{Filtering Principle}
\label{sec:filter}
We consider designing an exact solution to subtrajectory similarity search. 
A naive solution is enumerating all subtrajectories of the trajectories in 
$\database$ and computing the WED to the query. The time complexity is 
$O(\sum_{id=1}^N \size{P^{(id)}}^3 \cdot \size{Q})$. An improvement is achieved by the 
Smith-Waterman (SW) algorithm~\cite{SW}~\footnote{The SW here is slightly different 
from the standard one which performs local alignment. We adapt SW to our problem by 
changing the boundary condition of dynamic programming. 
\confversion{The pseudo-code is given in Appendix A of the extended version~\cite{extended-version}.}
\fullversion{See Appendix~\ref{appendix:pseduo-codes} for the pseudo-code.}}. 
It avoids enumerating subtrajectories and checks if 
$Q$ matches some $P_{i:j}$ in $O(\size{P} \cdot \size{Q})$ time (note that the 
threshold $\tau$ can be exploited for speed-up but it does not improve the time 
complexity), hence reducing the time complexity of processing a query to 
$O(\sum_{id=1}^N \size{P^{(id)}} \cdot \size{Q})$. However, this is still inefficient 
when the dataset is large or the trajectories are long. Our experiments show 
that it spends more than 30 minutes to answers a query on a dataset of 1 
million trajectories, using \netedr as distance function. 

Observing the inefficiency of the above baseline methods, we resort to indexing 
trajectories offline and answering the online query with a \emph{filter-and-verify} 
paradigm: by a filtering strategy, we first find a set of candidates, i.e., 
the data trajectories (along with the positions) that are probable to yield a 
match; and subsequently verify these candidates. Note that traditional filter-and-verify 
techniques for strings (e.g., $q$-grams and partition-based methods) are \emph{inefficient} 
or \emph{inapplicable} for WED on subtrajectories, as we have discussed in 
\S~\ref{sec:intro}. %

\subsection{{\Filtername}}\label{sec:sec:gsf}
The basic idea of our novel filtering principle, referred to as the 
\emph{\filtername}, is to choose a subsequence $Q'$ of the query $Q$ and derive 
a lower bound of WED using $Q'$. To guarantee to find all the answers to the 
query, we need a subsequence $Q'$ such that if the lower bound reaches $\tau$, then any 
trajectory $P$ having a subtrajectory match to $Q$ must contain at least one element 
(or one neighbor of the element) in $Q'$. To this end, we start with analyzing the effect of 
edit operations. 

Given two (non-empty) trajectories $P$ and $Q$ on $\Sigma$, consider converting $Q$ into $P$ with some edit operations.
If a symbol $q\in Q$ does not appear in $P$, $q$ must be substituted or deleted from $Q$.
For deletion, we need to pay a cost $\sub(q,\varepsilon)$.
To substitute $q$ with any $q'\in\Sigma\backslash\{q\}$, we need a cost $\sub(q,q')$.
Therefore, to substitute or delete $q$, the cost is at least
\begin{align*}
    \tilde c(q):=\min_{q'\in\SigmaEps\backslash\{q\}}\sub(q,q').
\end{align*}

\begin{exam}\label{example:idea}
Consider two trajectories $P=\texttt{BCD}$ and $Q=\texttt{ABC}$ on an alphabet $\Sigma=\{\texttt{A},\texttt{B},\texttt{C},\texttt{D}\}$. Assume the following cost

{
\vspace{-.5em}
\begin{center}
\begin{tabular}{|c|c|c|c|c|} \hline
    $\sub(\texttt{A},\texttt{A})$ &
    $\sub(\texttt{A},\texttt{B})$ & $\sub(\texttt{A},\texttt{C})$  & $\sub(\texttt{A},\texttt{D})$  & $\del(\texttt{A})$ \\ \hline
    0 & 5 & 3 & 6 & 4 \\ \hline
\end{tabular}    
\end{center}
\vspace{.3em}\,\hspace{-0.6em}
}%
Consider $q=\texttt{A}$ in $Q$. We see that \texttt{A} does not appear in $P$ and the minimum 
cost to delete or substitute this \texttt{A} turns out to be $\tilde c(\texttt{A})=\sub(\texttt{A},\texttt{C})=3$.
\end{exam}

Based on the cost $\tilde c(q)$, we can derive a lower bound.
For the general case of WED, an issue is that this lower bound can become loose if there exists 
$q'\in\SigmaEps\backslash\{q\}$ with a small cost; 
e.g., considering \edr, even if $q'\ne q$, $\sub(q,q')$ can be zero 
(when $d(q,q')\le\varepsilon$).
To address this issue and derive the filtering principle, we propose the concept of 
\emph{substitution neighbors}. 

\begin{defi}[Substitution Neighbors]
\label{def:sub-nei}
Given a symbol $q \in \Sigma$, the \emph{substitution neighbors} of $q$ are defined by
\begin{align}
    B(q):=\{b\in\Sigma\mid\sub(q,b)\le\eta\},
    \label{eq:sub-nei}
\end{align}
where $\eta\ge0$ is a cost threshold, depending on the cost function (we discuss the choice of $\eta$ below). 
For example, by setting $\eta=0$ in \edr, $B(q)$ is a set of 
vertices $b\in V$ such that $d(q,b)\le\varepsilon$ ($\Leftrightarrow\sub(q,b)=0$).
Note that $q\in B(q)$ always holds as $\sub(q,q)=0$.
Given a sequence $Q\in\Sigma^*$, we define the substitution neighbor of $Q$ by %
\begin{align}
    B(Q):=\bigcup_{q\in Q}B(q).
\end{align}
\end{defi}
Intuitively, $B(Q)$ is comprised of the vertices (or edges) that are either in $Q$ 
or too close to those in $Q$ to deliver significant cost. 
By considering the costs of the elements in $B(Q)$, we can obtain $c(Q)$, a lower 
bound of the cost of substituting or deleting all the elements in $B(Q)$, which 
leads to the following filtering principle\confversion{~\footnote{The proofs are given 
in Appendix B of the extended version~\cite{extended-version}.}}\fullversion{~\footnote{Please 
see Appendix~\ref{appendix:proofs} for the proof.}}.

\begin{theo}[\Filtername~Principle]\label{thm:gsf}
\revise{Given a subtrajectory $P' \substr P \in\Sigma^*$ and a query $Q\in\Sigma^*$, 
suppose a subsequence $Q'\subseq Q$ such that $P' \cap B(Q')=\emptyset$ and 
$c(Q'):=\sum_{q\in Q'}c(q)\ge\tau$ where
\begin{align}
    c(q):=\min_{q'\in\SigmaEps\backslash B(q)}\sub(q,q').
\end{align}
Then $\wed(P',Q) \geq \tau$.}
\end{theo}

We refer to a subsequence $Q'$ satisfying $c(Q')\ge\tau$ as a \emph{$\tau$-subsequence} of $Q$. 
The filtering principle states that if $P'$ does not share any element with $B(Q')$, 
where $Q'$ is a $\tau$-subsequence of $Q$, then it is guaranteed that $P'$ is not a result. 
Computing each $c(q), q \in Q'$is sublinear-time w.r.t. $\size{V}$ or $\size{E}$:
For \erp, the complexity is $O(\log\size{V})$ using a $k$d-tree. For other similarity functions in \S~\ref{sec:prelim:ged}, 
the complexity is $O(1)$, because $c(q)$ is $1$ for \edr, \lev, and \netedr, the smallest edge cost from $q$ for \neterp, and $\del(q)$ for \surs. 
Further, we make two remarks on this filtering principle:
\begin{itemize}
    \item The filtering is not limited to a specific cost function. Hence it can be 
    used for the general purpose.
    \item $Q'$ can be an arbitrary subsequence of $Q$ (we discuss the choice of $Q'$ 
    in \S~\ref{sec:filter:minknapsack}).
\end{itemize}
\revise{Candidates are only found in the trajectories that pass the filtering principle. 
Since we are going to look up an inverted index with the elements in $B(Q')$ to identify candidates 
(\S~\ref{sec:index}), we denote each candidate by a triplet $(id, j, i_q)$. $id$ is the trajectory 
ID. $j$ and $i_q$ are positions in $P^{(id)}$ and $Q$, respectively, at which the candidate 
is identified; i.e., $P_j^{(id)} \in B(Q_{i_q}), Q_{i_q} \in Q'$.}

\begin{exam}\label{example:gsf}
Suppose $\eta=0$ and the following cost matrix.

{
\vspace{-.5em}
\begin{center}
\small
\begin{tabular}{|c|c|c|c|c|c||c|} \hline
    $q$ &
    $\sub(q,\texttt{A})$ &
    $\sub(q,\texttt{B})$ &
    $\sub(q,\texttt{C})$ &
    $\sub(q,\texttt{D})$ &
    $\del(q)$ & $c(q)$\\ \hline
    $\texttt{A}$ & 0 & 5 & 3 & 6 & 4 & 3\\
    $\texttt{B}$ & 5 & 0 & 2 & 0 & 1 & 1\\
    $\texttt{C}$ & 3 & 2 & 0 & 5 & 3 & 2\\
    $\texttt{D}$ & 6 & 0 & 5 & 0 & 4 & 4\\ \hline
\end{tabular}
\end{center}
\vspace{-.5em}\,
}
Consider $Q=\texttt{ABC}$. 
By definition, we have
$B(\texttt{A})=\{\texttt{A}\}$,
$B(\texttt{B})=\{\texttt{B},\texttt{D}\}$,
$B(\texttt{C})=\{\texttt{C}\}$, and
$B(\texttt{D})=\{\texttt{B},\texttt{D}\}.$
Taking the minimum over $\SigmaEps\backslash B(q)$ for each row, we have $c(q)$ as in the table above. 
Consider $P^{(1)}=\texttt{BCDBCD}$,  $P^{(2)}=\texttt{DABCBA}$, and $P^{(3)}=\texttt{ABABAB}$. Consider $\tau=3$ and  $Q'=\texttt{A}\subseq Q$, which satisfies $c(Q')\ge\tau$.
As $P^{(1)}\cap Q'=\emptyset$, $P^{(1)}$ can be pruned  
(its subtrajectory closest to $Q$ is $\texttt{BC}$, where $\wed(\texttt{BC},Q)=\del(\texttt{A})=4\not<\tau$). 
\revise{Since $P^{(2)}$ and $P^{(3)}$ contains \texttt{A}, they pass the filter and 
generate candidates: $(P^{(2)}, 2, 1)$, $(P^{(2)}, 6, 1)$, $(P^{(3)}, 1, 1)$, $(P^{(3)}, 3, 1)$, and 
$(P^{(3)}, 5, 1)$. 
By verification, only $(P^{(2)}, 2, 1)$ yields a result \texttt{ABC} because $\wed(\texttt{ABC},Q)=0<\tau$. 
Other candidates are false positives (e.g., the closest subtrajectory to $Q$ in $P^{(3)}$ is \texttt{ABA}, 
where $\wed(\texttt{ABA},Q)=3\not<\tau$).
} 
\end{exam}

\myparagraph{Choice of $\eta$}
\begin{inparaenum} [(1)]
  \item For discrete cost functions (e.g., \lev and \edr), 
  we can use $\eta=0$, which excludes only symbols $a\in\Sigma^+$ with $sub(q,a)=0$, $q \in Q'$. 
  \item %
  For continuous cost functions (e.g., \erp), since the cost can be arbitrarily small, we need a 
  small positive number for $\eta$ to prevent the lower bound, $c(Q')$, becoming too loose (an 
  extreme case is that $c(Q) < \tau$, making the choice of $\tau$-subsequence impossible). We may 
  tune $\eta$ for the tightness of $c(Q)'$. 
  With increasing $\eta$, $c(Q')$ increases, leading to a 
  tighter lower bound; however, the number of symbols in $B(Q')$ also increases, which results 
  in a larger candidate set. %
  Setting $\eta$ to $\frac{\tau}{\size{Q}}$ guarantees that a $\tau$-subsequence can be found. 
  Our empirical study shows that a small positive number is good for continuous cost functions 
  (see \S~\ref{sec:experiments:setup} for experiment setting). 
\end{inparaenum}

\subsection{Finding Optimal {\large $\tau$}-Subsequence}
\label{sec:filter:minknapsack}
The \filtername~(Theorem~\ref{thm:gsf}) holds for any $Q'\subseq Q$ that satisfies 
$c(Q')\ge\tau$. %
To reduce computational cost in 
verification, we propose to choose a subsequence that
minimizes the number of candidates. This is formulated as a discrete optimization problem, as below.

According to the \filtername, $(P,T)\in\database$ such that $P\cap B(Q')\ne\emptyset$ 
will generate candidate trajectories.
Let $n(q)$ be the frequency of a symbol $q\in\Sigma$ that appears in $\mathcal{T}$. 
We note that the frequency is counted multiple times if $q$ occurs multiple times in a 
data trajectory, \revise{because we also record the positions $j$ and $i_q$ in a candidate.} 
The total number of symbols in $\database$ that intersects with $B(q)$ is $\sum_{b\in B(q)}n(q)$.
Therefore, we can formulate an optimization problem that minimizes the number of candidates as follows.
\begin{defi}[Minimum Candidate Problem]
\label{defi:MCP}
The minimum candidate problem \textsc{(MinCand)} is to find a subsequence $Q'\subseq Q$ 
defined by the following discrete optimization problem:
\begin{align}
    \min_{Q'\subseq Q} \sum_{q\in Q'}\sum_{b\in B(q)}n(b),\quad
    \text{\emph{subject to}}\quad \sum_{q\in Q'}c(q)\ge\tau.
    \label{eq:mincand-obj}
\end{align}
\end{defi}

\begin{exam}
Consider the same trajectories and cost matrix as Example~\ref{example:gsf}. 
The frequencies are $n(\texttt{A})=5$, $n(\texttt{B})=7$, $n(\texttt{C})=3$, and $n(\texttt{D})=3$. 
There are seven subsequences of $Q=\texttt{ABC}$, namely \texttt{A}, \texttt{B}, \texttt{C}, \texttt{AB}, \texttt{AC}, \texttt{BC}, and \texttt{ABC}.
Among them, \texttt{A}, \texttt{AB}, \texttt{AC}, \texttt{BC}, and \texttt{ABC} satisfy the constraint $\sum_{q\in Q'}c(q)\ge\tau=3$.
Evaluating the objective function for each subsequence, we obtain:

{
\vspace{-.5em}
\begin{center}
\begin{tabular}{|c|c|c|c|c|c|} \hline
    $Q'$ &
    $\texttt{A}$ &
    $\texttt{AB}$ &
    $\texttt{AC}$ &
    $\texttt{BC}$ &
    $\texttt{ABC}$ \\ \hline
    Obj. & 5 & 15 & 8 & 13 & 18\\ \hline
\end{tabular}
\end{center}
\vspace{-.5em}\,\hspace{-.5em}
}%
Hence, $Q'=\texttt{A}$ is the optimal solution
(note: as $B(\texttt{B})=\{\texttt{B},\texttt{D}\}$, we have $\sum_{b\in B(\texttt{B})}n(b)=n(\texttt{B})+n(\texttt{D})=10$). 
\end{exam}

\myparagraph{Remark} 
Given symbols $q$ and $q'$ in $Q$, if $B(q)\cap B(q')\ne\emptyset$, 
say $q''$ is a common element of $B(q)$ and $B(q')$, then Eq.(\ref{eq:mincand-obj}) 
counts trajectories that travel on $q''$ twice.
\revise{The elements counted multiple times are treated distinctly as they 
correspond to different candidates (distinct $i_q$ and $i_{q'}$). 
To see the formulation does not violate the correctness of the search algorithm, 
there are $\size{Q'}$ elements in $Q'$, each $q \in Q'$ has 
$\size{B(q)}$ substitution neighbors, and each neighbor $b$ generates $n(b)$ 
candidates (i.e., the number of $(id, j)$ pairs is $n(b)$ in $\mathcal{T}$). 
Besides, there is no duplicate among these candidates due to distinct $(id, j)$ 
and $i_q$. Hence the objective in Eq.(\ref{eq:mincand-obj}) is exactly the 
candidate size.} 
Next we discuss the computational aspects of the \textsc{MinCand} problem -- 
it is NP-hard but polynomial-time 2-approximation is available.
An observation is that \textsc{MinCand} is similar to the 0-1 knapsack problem. %
In fact, it can be reduced from the \emph{Minimum Knapsack Problem} (MKP)~\cite{min-knapsack}, defined as follows.
\begin{align}
    \min_{S \subseq [\![K]\!]} \sum_{k \in S} W_k,\quad
    \text{subject to}\quad \sum_{k \in S} V_k \ge D. 
    \label{eq:min-knapsack}
\end{align}
$K$ is the number of items; 
$W_k$ is the weight of an item and $V_k$ is its value. The goal is to select a minimum weight subset of items 
$S \subseq [\![K]\!]$, such that the total value is no less than a demand $D$. Hence we have 

\begin{prop}
\label{theo:np-hard}
\textsc{MinCand} is NP-hard.
\end{prop}

Seeing the NP-hardness, we employ an approximation algorithm (Algorithm~\ref{algo:primaldual}) based on \cite{min-knapsack} 
(\textsc{MinCand} can be reduced to MKP and thus we can use the algorithm for MKP, see Proposition~\ref{theo:approx-ratio}). 
\revise{
In brief, this algorithm starts with an empty subsequence $Q'$, and greedily adds an item $q^*$ that has the minimum $v_q$ value (Lines~4--5) to $Q'$. 
Intuitively, if $N_q/c(q)$ is small, the item would be worth choosing because we want to choose one with small $N_q$ and large $c(q)$. 
Following the justification in \cite{min-knapsack}, we extend this idea with a slight modification, and 
use $(N_q-w_q)/\min(c(q),\tau-c(Q'))$ as $v_q$, where 
the $w_q$ variables (Line~6) are related to the dual problem of Eq.~(\ref{eq:min-knapsack}). %
We stop this procedure when the constraint in Eq.~(\ref{eq:mincand-obj}) (i.e., $\tau\le c(Q'):=\sum_{(q,i_q)\in Q'}c(q)$) is satisfied.
At Line~7, we also record $i_q$, which is the position of $q$ in $Q$. This information is carried when candidates are generated. %
}

\begin{algorithm}[t]
\caption{$\textsf{MinCand}(Q, n, c, \tau)$} %
\label{algo:primaldual}
\small
$N_q\leftarrow\sum_{b\in B(q)}n(b)\quad (\forall q\in Q)$\tcp*{Trajectory freq.}
$Q'\leftarrow \emptyset$; \, $w_q\leftarrow 0\quad (\forall q\in Q)$\tcp*{Initialize}
\While(\hspace{9.2em}\tcp*[h]{Constraint~\eqref{eq:mincand-obj}}){$\tau>c(Q')$}{
    \revise{$v_q\leftarrow (N_q-w_q)/\min\{c(q), \tau-c(Q')\}\quad \forall q\in Q\backslash Q'$\;}
    $q^*\leftarrow\text{argmin}_{q\in Q\backslash Q'}\{v_q\}$\tcp*{Choose greedily}
    \revise{$w_q\leftarrow w_q + \min\{c(q), \tau-c(Q')\} \cdot v_{q^*}\quad\forall q\in Q\backslash Q'$\;}
    $Q'\leftarrow Q'\cup\{(q^*,i_{q^*})\}$\tcp*{$i_{q^*}$:position of $q^*$ in $Q$}
}
\Return{$Q'$}
\end{algorithm}

\begin{exam}
Suppose that $Q=\texttt{ABCD}$, $c=[1,2,3,4]$ and $N=[5,2,9,8]$.
If $\tau=4$, we have $v=[5,\underline{1},3,2]$. Hence, we add the second item, \texttt{B}, and its position, 2, to $Q'$.
Then we update $w=[1,2,3,4]$ and $\tau-c(Q')= 4-c(\texttt{B})=2$.
In the next iteration, we have $v=[4/1,\texttt{-},6/2,4/2]$ and we add the forth item, $(\texttt{D},4)$ to $Q'$.
This results in $\tau-c(Q')=4-c(\texttt{B})-c(\texttt{D})=-2$ and we stop the iteration and obtain $Q'=\{(\texttt{B},2), (\texttt{D},4)\}$.
Although this $Q'$ is not the optimal one $Q^*=\{(\texttt{D},4)\}$, we have a good approximation (10/8=25\% loss compared to the optimal). %
\end{exam}

\myparagraph{Algorithm Property}
Algorithm~\ref{algo:primaldual} runs in $O(|Q|^2)$ time. Further, the following statement holds. %
\begin{prop} \label{theo:approx-ratio}
 Let $f^*$ be the optimal objective value of Eq.~(\ref{eq:min-knapsack}).
 The approximation ratio of Algorithm~\ref{algo:primaldual} is 2, i.e., the approximated objective value is not greater than $2f^*$.
\end{prop}
For a special case, the following stronger result holds (\edr, \lev, and \netedr satisfy this property).
\begin{prop} \label{theo:global-optim}
If $c(q)$ is a constant function, i.e., $c(q)=c'$, Algorithm~\ref{algo:primaldual} returns the optimal solution of \textsc{MinCand}.
\end{prop}

Solving \textsc{MinCand} is similar to finding best substrings (incl. $q$-grams) for string similarity problems~\cite{DBLP:conf/vldb/LiWY07,DBLP:conf/sigmod/YangWL08,DBLP:journals/tkde/WangQXLS13,DBLP:journals/tods/Qin0XLLW13,DBLP:journals/tods/LiDF13}. 
The main differences are: 
\begin{inparaenum} [(1)]
  \item They mainly target Levenshtein distance on entire strings, while we cope with WED on substrings. 
  \item They resort to either heuristics~\cite{DBLP:journals/tods/Qin0XLLW13,DBLP:journals/tods/LiDF13} 
  or an offline constructed 
  dictionary~\cite{DBLP:conf/vldb/LiWY07,DBLP:conf/sigmod/YangWL08,DBLP:journals/tkde/WangQXLS13} without 
  performance guarantee on candidate size, while we model this as a discrete optimization problem solved 
  by a 2-approximation algorithm. 
\end{inparaenum}
 
\section{Indexing and Search Algorithm}
\label{sec:index}

\subsection{Indexing}

Our indexing method employs inverted index~\cite{manning}, which is widely used for keyword search 
and also used to deal with trajectory similarity search (e.g., \cite{SIGIR18-Torch,ICDE18-DISON}).
We store data trajectories in the postings list (denoted by $L_q$) of each symbol $q\in\Sigma$.
A record in $L_q$ is in the form of $(id, j)$, where $id$ is the ID of a trajectory that passes 
$q$ and $j$ is its position, i.e., $P^{(id)}_j=q$. 
We can update the index by appending a new record to the corresponding postings list.

\subsection{Search Algorithm} \label{sec:index:search-algorithm}
We propose an algorithm for subtrajectory similarity search problem based on the \filtername~in \S~\ref{sec:filter}.
Algorithm~\ref{algo:s3} shows a skeleton of the algorithm.

Given a query $Q$, we first generate candidates based on Theorem~\ref{thm:gsf}.
This states that for any subsequence $Q'$ satisfying $c(Q')\ge\tau$, trajectories not included 
$\mathcal{C}=\cup_{b\in B(Q')}L_{b}$ can be safely pruned.
To minimize the size of this candidate set $\mathcal{C}$, we solve the \textsc{MinCand} using Algorithm~\ref{algo:primaldual} (at Line~1 of Algorithm~\ref{algo:s3}).
We iterate through each $(q,i_q) \in Q'$ and look up the postings list of $b \in B(q)$. 
At Line~6, the ID of the data trajectory that contains $b\in B(q)$ is added to the 
candidate set. We also include the corresponding positions in $P$ and $Q$ (denoted 
by $j$ and $i_q$, respectively).
They are used to speed up the verification (\S~\ref{sec:verify}).
After the candidates are obtained, we verify whether each of them truly matches 
the query $Q$.

\begin{algorithm}[t]
 \small
 \caption{$\textsf{SubtrajSimSearch}(Q, \database, \wed, \tau)$}
 \label{algo:s3}
 \SetKwInOut{Input}{input}
 \SetKwInOut{Output}{output} \DontPrintSemicolon
 \Input{Query: $Q$;
 Database: $\database$;
 Similarity function: $\wed$;
 Similarity threshold: $\tau$}
 $Q'\leftarrow \tsf{MinCand}(Q, n, c, \tau)$\tcp*{Optimize $\tau$-subsequence}
 $\mathcal{C}\leftarrow\emptyset$\;
 \For{$(q,i_q)\in Q'$}{
   \For{$b\in B(q)$}{
     \For{$(id,j)\in L_{b}$}{
       $\mathcal{C}\leftarrow\mathcal{C}\cup\{(id, j, i_q)\}$
     }
   }
 }
 $\mathcal{A}\leftarrow\tsf{Verify}(\mathcal{C},Q,\tau)$\tcp*{See Algorithm~\ref{algo:verify} in \S~\ref{sec:verify}}
 \Return{$\mathcal{A}$}
\end{algorithm}

\myparagraph{Incorporating spatial/road network indexing}
Despite focusing on the general case of WED, it is noteworthy to mention that 
we can improve the query processing by indexing spatial information for a specific 
similarity function and regarding the index as a blackbox without needing to modify our 
algorithm. For similarity functions that involves Euclidean distance, we 
may index the coordinates of the vertices $V$ using a spatial index, such as a $k$d-tree or an R-tree, 
so as to quickly compute $B(q)$ by retrieving the symbols within a range to $q$. %
For similarity functions involving shortest path distance (e.g., \netedr~and \neterp), 
we may use the \emph{hub-labeling index}~\cite{Abraham:2012:HHL:2404160.2404164,Akiba:2013:FES:2463676.2465315} 
to compute shortest path distance to get $\sub(v,v')$. 
\confversion{We provide a running example in the extended version~\cite{extended-version}.}

\fullversion{
\begin{exam}
Figure~\ref{fig:index-structure} shows an example of indexing and query processing. 
When building index, given a trajectory $P^1=v_2v_3v_5$, we store $(id, j)=(1, 1)$ to the 
postings list of $v_2$, $(1, 2)$ to $v_3$, and $(1, 3)$ to $v_5$. Suppose the similarity 
function is \edr, and we have a query $Q$ whose $Q'$ returned by \textsc{MinCand} is 
$\set{(v_1, 3)}$. By utilizing the spatial index for range query, we have $B(q) = \set{v_1, 
v_2}, q \in Q'$. By looking up the postings lists of $v_1$ and $v_2$, we find a candidate 
$(1, 1, 3)$ to be verified.
\end{exam}

\begin{figure}[ht]
    \centering
    \includegraphics[width=\linewidth]{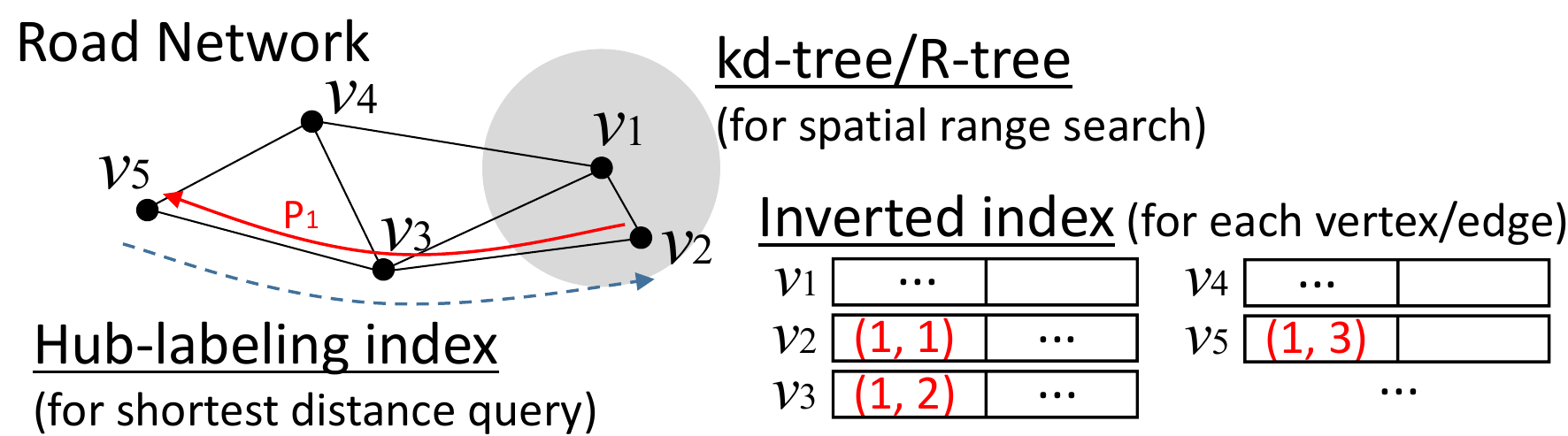}
    \caption{Index structure.}
    \label{fig:index-structure}
\end{figure}
}
\subsection{Filtering with Temporal Information}
\label{sec:index:extension}
As mentioned in \S~\ref{sec:prelim:problem}, we can treat any kind of temporal constraints as postprocessing. 
For interval constraints, such as $[T_i,T_j]\subseteq I$ or $[T_i,T_j]\cap I\ne\emptyset$, we can \emph{prune candidates} before verification as follows.
For each candidate trajectory $(P^{(id)},T^{(id)})$ of length $n$, we check its first and last timestamps (i.e., $I^{(id)}:=[T_1^{(id)}, T_{n}^{(id)}]$).
Given a query time interval $I$, if $I^{(id)}\cap I=\emptyset$, then we have $[T_i,T_j]\not\subseteq I$ and $[T_i,T_j]\cap I\ne\emptyset$; we can safely prune this candidate. 
Furthermore, depending on the application, we may sort the records in each postings list by their temporal information such as departure time (i.e., $T^{(id)}_1$) or maximum speed (i.e., $\max_{1\le t\le |T^{(id)}|-1} w(P_t^{(id)})/(T^{(id)}_{t+1}-T^{(id)}_t)$, where $w(e)$ is the distance of an edge $e$). This allows us to \emph{generate candidates} with binary search on postings lists, hence to avoid those violating the temporal constraint.

\section{Verification}
\label{sec:verify}
\begin{figure}
    \centering
    \includegraphics[scale=0.7,trim=0 1em 0 1em]{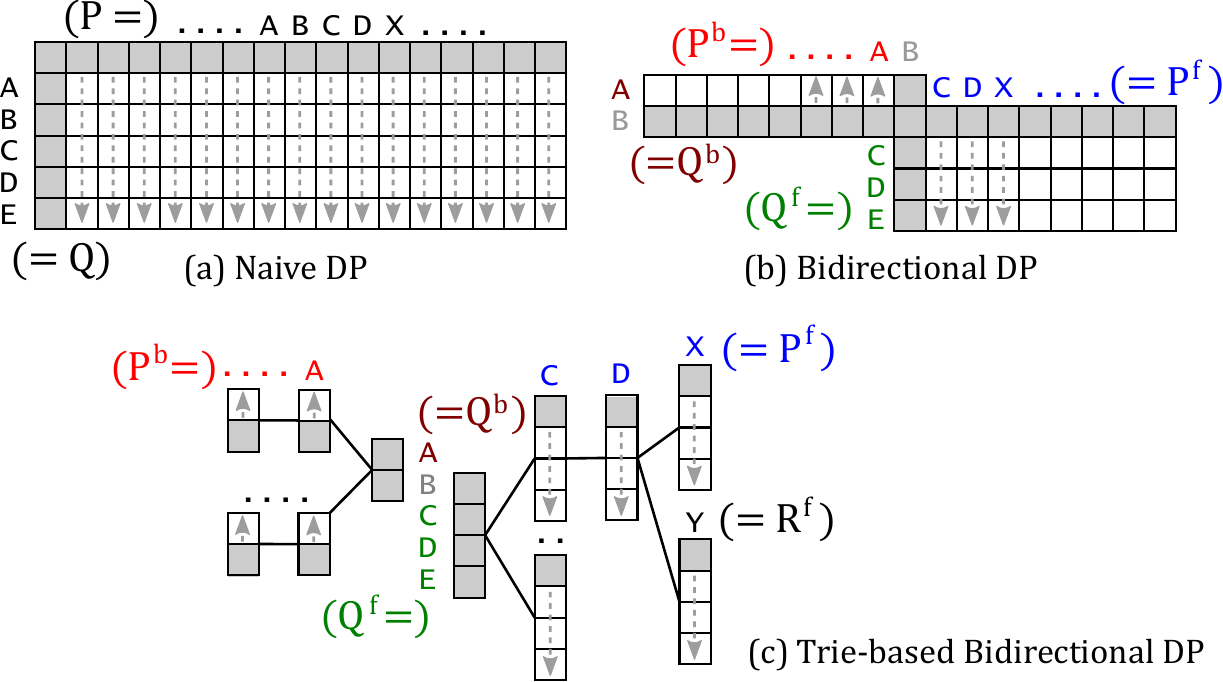}
    \caption{Comparison of verification methods 
    ($P=$ {\color{red}{\tsf{A}}}{\color{gray}{\tsf{B}}}{\color{blue}{\tsf{CDX...}}}, $Q=$ {\color{brown}{\tsf{A}}}{\color{gray}{\tsf{B}}}{\color{Green}{\tsf{CDE}}}; colors show partition).}
    \label{fig:verification-methods}
\end{figure}

The generated candidates usually include many false positives; 
therefore, we need to verify them efficiently to obtain the answer.
Existing methods for whole matching similarity search (e.g., 
\cite{SIGIR18-Torch,ICDE18-DISON}) computes the similarity between 
$P$ and $Q$ with dynamic programming (DP) for each candidate.
Similarly, in our subtrajectory search setting, we can 
compute the similarity by sequentially filling a $|P| \cdot |Q|$ 
matrix based on the recursive definition of WED, which is referred 
to as the Smith-Waterman (SW) algorithm. The time complexity is 
$O(|P|\cdot|Q|)$, as shown in Figure~\ref{fig:verification-methods}(a).
This naive SW algorithm includes the following redundant computation:
\begin{inparaenum} [(1)]
  \item Although a subtrajectory of $P$ similar to $Q$ can be a small 
  part of $P$, the SW algorithm computes DP matrix for the entire $P$.
  \item If two trajectories share a subtrajectory, the DP matrices have 
  common values, but the SW algorithm does not exploit this property.
\end{inparaenum}

\myparagraph{Main Idea}
We reduce redundant computation as follows. %
\begin{inparaenum} [(1)]
  \item \textbf{Local verification}: Given a candidate $(id,j,i_q)$, the 
  subtrajectory similar to $Q$ are located around the position $j$ of $P$; 
  hence we only need to run DP around $j$.
  \item \textbf{Trie-based caching}: We share computation for common 
  subtrajecories by exploiting the sparsity of a road network: Although the 
  alphabet $\Sigma$ is large, the possible previous/next symbols of $P_j$ 
  (i.e., $P_{j-1}$ and $P_{j+1}$) 
  are limited because trajectories move along physically connected vertices 
  (edges) in a road network; hence we can efficiently cache the columns of DP matrices.
\end{inparaenum}

\subsection{Local Verification}
\label{sec:bidirectional-search}
The filtering phase gives $(id, j, i_q)$, where $j$ is a position such that $P^{(id)}_j$ matches 
$Q_{i_q}$ or its substitution neighbor. 
The local verification is to check if there exists a subtrajectory $P_{s:t}^{(id)}$ such that 
$\wed(P_{s:t}^{(id)},Q) < \tau$, where $s \le j \le t$. 
Our idea is to run the DP computation from $j$ \emph{bidirectionally}. To guarantee that we will not 
miss any similarity search result after verifying all the candidates, we have the following lemma.

\begin{lemm}
  \label{theo:partition-dp}
  Given a subtrajectory $P_{s:t}^{(id)}$ of $P$ such that $\wed(P_{s:t}^{(id)},$ $Q)$ $< \tau$, and the 
  set of candidates $\mathcal{C}$ identified by subsequence filtering, 
  there exists $(id, j, i_q) \in \mathcal{C}$ such that $s \le j \le t$ and 
  \begin{align}
  \wed(P_{s:t}^{(id)},Q)=\;&\wed(P_{s:j-1}^{(id)},Q_{1:i_q-1})+\sub(P_j^{(id)},Q_{i_q}) \nonumber \\
  &+\wed(P_{j+1:t}^{(id)},Q_{i_q+1:|Q|}).
  \label{eq:wed-decomp}
  \end{align}
\end{lemm}

This lemma suggests that for every subtrajectory that satisfies the WED constraint, we can always 
find a position $j$ in the candidates identified by subsequence filtering, such that $P_j^{(id)}$ is 
aligned to $Q_{i_q}$ in the optimal alignment that yields the WED. Thus, we can verify from $j$ to obtain 
the similarity search result. Specifically, by Eq.~\eqref{eq:wed-decomp}, we partition $P_{s:t}^{(id)}$ 
into three parts at $j$ %
and compute WED bidirectionally~\cite{DBLP:conf/spire/HyyroN03}. 
From $j$, we run two DPs: a \textbf{b}ackward one (i.e., from the end of strings to the start) for 
$\wed(P_{s:j-1}^{(id)},Q_{1:i_q-1})$ and a \textbf{f}orward one for 
$\wed(P_{j+1:t}^{(id)},Q_{i_q+1:|Q|})$. By iterating $s$ from $j$ to $1$ and $t$ from $j$ to 
$\size{P^{(id)}}$ in the two DPs, respectively, we are able to find all $P_{s:t}^{(id)}$ such that 
$s \le j \le t$ and $\wed(P_{s:t}^{(id)}, Q) < \tau$. The above step is conducted for each 
$(id, j, i_q)$ in the candidate set to obtain all the similarity search results. For ease of 
exposition, in the rest of this section, we omit the superscript $(id)$ from $P$, and we use 
$P^b$ to denote $P_{s:j-1}$ and $P^f$ to denote $P_{j+1:t}$. 

\begin{exam}
\label{exam:bidp}
Consider a trajectory $P=...\textsf{ABCDX}...$ and a query $Q=\textsf{ABCDE}$. 
Assume a $\tau$-subsequence is $Q'=\{\textsf{B}\}$ and there is only one candidate 
whose $j = 2$ and $i_q = 2$. 
As shown in Figure~\ref{fig:verification-methods}(b), we partition $P$ into  $(P^b,P_j,P^f)=(...\textsf{A},\textsf{B},\textsf{CDX}...)$ and $Q$ into  $(Q^b,Q_{i_q},Q^f)=(\textsf{A},\textsf{B},\textsf{CDE})$.
We can find all subtrajectories $P_{s:t}$ such that $s \le 2 \le t$ and $\wed(P_{s:t}, Q) < \tau$ 
by computing $\sub(\textsf{B},\textsf{B})+\wed(...\textsf{A},\textsf{A})+\wed(\textsf{CDX}...,\textsf{CDE})$. 
\end{exam}

\noindent\textbf{Early Termination.}
For each direction $d\in\{b,f\}$, we can terminate the computation before reaching the end of 
$P^d$. Given a position $0 \leq k \leq \size{P^d}$ ($k = 0$ for the case of an empty string), 
by the definition of WED, we have the following lower bound of the WED between $P^d$ and $Q$: 
\begin{align}
  LB_k^d:=\min_{0\le j\le|Q|}\{\wed(P_{1:k}^d,Q_{1:j})\}\le\wed(P^d,Q). 
  \label{eq:lower-bound}
\end{align}
If the lower bound for any $k$ reaches $\tau$, we can safely terminate the DP computation of 
$P^d$. 

\subsection{Caching with Bidirectional Trie}
\label{sec:verify:trie}

The local verification still involves redundant computation when we verify multiple candidates.
We begin with an example.

\begin{exam}
\label{exam:common-prefix}
(Continuing from Example~\ref{exam:bidp})
Consider another trajectory $R=...\textsf{ABCDY}...$ identified as a candidate via $\textsf{B}$. 
In the local verification, we need to compute both 
$\wed(P^f,Q^f)$ and $\wed(R^f,Q^f)$, where $Q^f=\textsf{CDE}$, $P^f=\textsf{CDX}...$, and $R^f=\textsf{CDY}...$.
Hence, when computing $\wed(P^f,Q^f)$ and $\wed(R^f,Q^f)$, the first two columns of 
the DP matrices share the same values 
because $P^f$ and $R^f$ has a common prefix $\textsf{CD}$.
This indicates that we can reduce the computation by caching these columns.
\end{exam}

In general, given a vertex/edge in the road network, the number of possible next vertices/edges are very small 
(typically, three) compared to the alphabet size  because of the structure of the road network.
This implies that the candidate subtrajectories starting from $Q_{i_q}$ tend to share a prefix; therefore, 
we expect that the caching strategy will improve the efficiency.

To efficiently cache the DP columns of common prefixes, we employ a trie-based data structure as shown in Figure~\ref{fig:verification-methods}(c).
We describe how this trie works using a running example.

\begin{exam}
(Continuing from Example~\ref{exam:common-prefix})
After computing $\wed(P^f,Q^f)$, we have three columns corresponding to \textsf{C}, \textsf{D}, and \textsf{X}, as in Figure~\ref{fig:verification-methods}(c).
When computing $\wed(R^f,Q^f)$, we can reuse the two columns for \textsf{C} and \textsf{D} that are cached in the trie.
\end{exam}

As candidates tend to have common prefixes as discussed above, we expect that the cache miss rate is 
low %
and the verification gets faster. 
A trie is built for each direction (hence called a bidirectional trie) and each symbol in the 
$\tau$-subsequence of $Q$. So there are $2\size{Q'}$ tries. 

\subsection{Verification Algorithm}
Algorithms~\ref{algo:verify}--\ref{algo:one-step-DP} summarize our verification algorithm.
First, for each direction $\{b,f\}$ and $q\in Q'$ where $Q'$ is a $\tau$-subsequence of $Q$, empty tries are initialized (each trie is denoted by 
$\mathcal{T}_{i_q}^{b}$ or $\mathcal{T}_{i_q}^{f}$, where $i_q$ indicates the candidate position in $Q$).
Then, for each candidate $(id,j,i_q)\in\mathcal{C}$, \textsf{VerifyCandidate} (Algorithm~\ref{algo:verify-candidate}) is applied and the results are stored in $\mathcal{A}$.

In \textsf{VerifyCandidate}, the data trajectory $P$ and query $Q$ are partitioned into three parts. 
Then, for each direction $d\in\{b,f\}$, \textsf{AllPrefixWED} is called to compute an array $E^d$, 
whose $k$-th element is the WED between $Q^d$ and the $k$-th prefix of $P^d$, i.e., $E_k^d=\wed(P^d_{1:k},Q^d)$.
We can use a tighter threshold $\tau':=\tau-\sub(q,b)$ instead of the original $\tau$ because of Eq.~\eqref{eq:wed-decomp}; this $\tau'$ is used for early termination.
Finally, all the subtrajectories that satisfy Eq.~\eqref{eq:wed-decomp} $<\tau$ are added to the result set.

In \textsf{AllPrefixWED} (Algorithm~\ref{algo:all-prefix-wed}), we compute the array $E^d$.
A symbol $c$ in a given trajectory $P^d$ is processed one by one; if a child node corresponding to $c$ at the current trie is found (Line~3), we skip computing the corresponding DP column; otherwise, we create a new child (Line~5) and compute the DP column (Line~6) using the \textsf{StepDP} procedure (Algorithm~\ref{algo:one-step-DP}), a standard DP that computes a new column based on the previous column. $A^{(x)}$ represents a DP column cached in the trie node $x$.
By Eq.~\eqref{eq:lower-bound}, if the lower bound $LB_k^d$ exceeds a given threshold (Line~7), we can safely terminate the DP computation for $P^d$.
Finally, the value $A_{|Q_d|}^{(x)}=\wed(P^d_{1:k},Q^d)$ is stored to $E_k^d$ (Line~9).

\begin{algorithm}[t]
 \small
 \caption{$\textsf{Verify}(\mathcal{C}, Q, \tau)$}
 \label{algo:verify}
 \SetKwInOut{Input}{input}
 \Input{Candidates $\mathcal{C}$; Query $Q$; Threshold $\tau$}
 \For{$(q,i_q)\in Q'$}{
   $\trie_{i_q}^f\!, \trie_{i_q}^b\!\leftarrow\text{Empty tries}$
 } 
 \For{$(id, j, i_q)\in\mathcal{C}$}{
    $\mathcal{A}\leftarrow\mathcal{A}\cup\tsf{VerifyCandidate}(Q, (id,j,i_q), \tau, \trie_{i_q}^f, \trie_{i_q}^b)$
  }
 \Return{$\mathcal{A}$}
\end{algorithm}

\begin{algorithm}[t]
 \small
 \caption{$\textsf{VerifyCandidate}(Q, (id, j, i_q), \tau, \mathcal{T}^f, \mathcal{T}^b)$}
 \label{algo:verify-candidate}
 \SetKwInOut{Input}{input}
 \SetKwInOut{Output}{output} \DontPrintSemicolon
 \Input{Query: $Q\in\Sigma^*$;
 Candidate: $(id,j,i_q)$;
 Threshold: $\tau$;
 Forward/Backward tries: $\mathcal{T}^f, \mathcal{T}^b$}
 $P\leftarrow \tsf{accessTrajectory}(id)$\;
 $(P^b,b,P^f)\leftarrow(P_{1:(j-1)},P_j,P_{(j+1):|P|})$\tcp*{Partition}
 $(Q^b,q,Q^f)\leftarrow(Q_{1:(i_q-1)},Q_{i_q},Q_{(i_q+1):|Q|})$\tcp*{Partition}
 $E^b\leftarrow \tsf{AllPrefixWED}(Q^b,P^b,\tau',\mathcal{T}^b\!.\textsf{root})$\;%
 $E^f\leftarrow \tsf{AllPrefixWED}(Q^f,P^f,\tau',\mathcal{T}^f\!.\textsf{root})$\;%
 \For{$(s,t)$ \textbf{\emph{such that}} $\sub(q,b)+ E^b_s+ E^f_t<\tau$}{$\mathcal{S}\leftarrow \mathcal{S}\cup\{(id,s,t)\}$\tcp*{Initialized as $\mathcal{S}=\emptyset$}}
 \Return{$\mathcal{S}$}\tcp*{All subtrajectories $P_{s:t}$ that matches $Q$}
\end{algorithm}

\begin{algorithm}[t]
 \small
 \caption{$\textsf{AllPrefixWED}(Q^d, P^d, \tau, x)$}
 \label{algo:all-prefix-wed}
 \SetKwInOut{Input}{input}
 \SetKwInOut{Output}{output}
 \SetKwBlock{Repeat}{repeat}{}
 \DontPrintSemicolon
 \Input{Query $Q^d$; Trajectory $P^d$; Trie node $x$; ($d\in\{f,b\})$}
 \Output{WED between $Q^d$ and $P^d_{1:k}$ $(\forall k)$}
 $E^d\leftarrow \text{Empty array}$\tcp*{$E_k^d$ means $\wed(Q^d,P^d_{1:k})$}
 \For{$k$  \textbf{\emph{in}}  $1..|P^d|$}{
   $c:=P^d_k$;\hspace{5pt}
   $x_\textsf{parent}\leftarrow x$; \hspace{5pt} $x\leftarrow x_\textsf{parent}.\tsf{findChild}(c)$\;
   \If{$x$ \emph{\textsf{not found}}}{
     $x\leftarrow x_\textsf{parent}.\tsf{createChild}(c)$\tcp*{New child}
     $A^{(x)}\leftarrow\textsf{StepDP}(Q^d, c, A^{(x_\textsf{parent})})$ \tcp*{Fill DP column}
   }

   \If{$\tau\le\min_{0\le j\le |Q|} A_j^{(x)} \; (=LB_k^d)$}{\textbf{break}\tcp*{Early Termination (Sec.\ref{sec:bidirectional-search})}
 }
 $E_k^d\leftarrow A^{(x)}_{|Q^d|}$\tcp*{$\wed(Q^d,P^d_{1:k})$}
} %
\Return{$E^d$}
\end{algorithm}

\begin{algorithm}[t]
 \small
 \caption{$\textsf{StepDP}(Q^d, p, A)$}
 \label{algo:one-step-DP}
 \SetKwInOut{Input}{input}
 \SetKwInOut{Output}{output} \DontPrintSemicolon
 \Input{Query $Q^d$; Next symbol $p\in\Sigma$; DP array $A_{0:\size{Q^d}}$}
 $B\leftarrow$ Array of length $\size{Q^d}+1$; \hspace{5pt}
 $B_0\leftarrow A_0+\del(p)$\;
 \For{$j$ \emph{\textbf{in}} $1..\size{Q^d}$}{
     {\small\!$B_j\!\leftarrow\!\min\{A_{j-1}\!+\!\sub(p, Q^d_j), A_j\!+\!\del(p), B_{j-1}\!+\!\ins(Q^d_j)\}$}
 }
 \Return{$B$}
\end{algorithm}

\section{Experiments}
\label{sec:experiments}
\confversion{We report the most important experiments here. Please see \cite{extended-version} 
for detailed setup and additional experiments. 
}

\subsection{Settings}
\label{sec:experiments:setup}
Evaluation was conducted on the following datasets: \underline{\beijing} (T-drive) 
\cite{T-drive}, \underline{\porto}~\cite{PortoDatset-ECML},  \fullversion{\revise{\underline{\singapore}~\cite{DBLP:journals/pvldb/SongSZZ14},}} and \underline{\sanfran}~\cite{Brinkhoff}.
For \beijing~and \porto, we conducted map matching \cite{HM4} to obtain network-constrained representation.
\sanfran~is a large synthesized dataset by the moving object generator \cite{Brinkhoff} with the 
San~Francisco road network.
The statistics after preprocessing are presented in Table~\ref{tab:dataset-stats}.
\begin{table}[t]
    \centering
    \caption{Dataset statistics.}
    \label{tab:dataset-stats}
    \small
    \begin{tabular}{c|c|c|c|c} \hline
     Dataset     & \# Trajectories   & Avg. Length & $|V|$ & $|E|$ \\ \hline
     \beijing    & 786,801  & 101    & 86,484 & 171,135 \\
     \porto      & 1,701,238&  81    & 75,265 & 135,133 \\
     \fullversion{\revise{\singapore}  & \revise{287,524}  & \revise{262}    & \revise{18,127} & \revise{48,236} \\}
     \sanfran & 11,505,922 & 101 & 175,343 & 223,606 \\ \hline
    \end{tabular}
    \vspace{-1em}
\end{table}

Our method consists of the (optimized) \filtername~with the bidirectional trie (BT) verification (referred to as \underline{\tsf{\GSF-BT}}).
We also consider \underline{\tsf{\GSF-SW}}, where BT is replaced by the Smith-Waterman (SW) algorithm 
for verification. Further, we compare with the following baselines. 

\myparagraph{DISON}
\revise{\tsf{DISON}~\cite{ICDE18-DISON} is a whole matching method for \lcrs. 
We adapted it to our problem. 
Since the early termination technique in \cite{ICDE18-DISON} does not work here, we used SW 
(\underline{\tsf{DISON-SW}}) or BT verification (\underline{\tsf{DISON-BT}}).} 

\myparagraph{Torch}
\revise{\tsf{Torch}~\cite{SIGIR18-Torch} is a whole matching method that supports several similarity functions. We 
adapted it to our problem. %
We equipped it with SW (\underline{\tsf{Torch-SW}}) and BT (\underline{\tsf{Torch-BT}}) for 
verification. The upper bounding technique~\cite{SIGIR18-Torch} prior to verification was developed for 
\lors and does not apply here.}

\myparagraph{DITA}
\underline{\textsf{DITA}}~\cite{SIGMOD18-DITA} is a whole matching method developed for \dtw and can be adapted for other 
functions. 
We modified its pivoting method to fit \WED. %
Since \textsf{DITA} does not support subtrajectory search, we enumerated all subtrajectories and indexed them. 
\revise{Note the enumeration is done offline and not counted towards query processing time.} 
We used SW instead of the double-direction verification (DDV)~\cite{SIGMOD18-DITA} 
because DDV works for \dtw but does not improve upon SW for WED.

\myparagraph{$q$-gram indexing for \edr}
\underline{\tsf{$q$-gram}} indexing was proposed in \cite{EDR} 
for whole matching under the \edr based on the fact that if there are less than $\max\{|P|,|Q|\}-q+1-\tau q$ common $q$-grams 
between $P$ and $Q$, then $\edr(P,Q)>\tau$. We customized their method to support subtrajectory search under \edr. 
We set $q=3$. SW was used for verification. 

\myparagraph{Indexing for \erp}
\underline{\tsf{ERP-index}} was proposed in \cite{ERP}. It employs lower bounding and triangle inequality. Given a sequence $P$ 
of coordinates, we indexed the sum of all coordinates, $sum(P)\in\mathbb{R}^2$, in a spatial index (we used $k$d-tree). Given a 
query sequence $Q$, $\|sum(P)-sum(Q)\|$ gives a lower bound.
We enumerated and indexed all subtrajectories. SW was used for verification. 

\myparagraph{Smith-Waterman (SW)}
The SW algorithm~\cite{SW} is a non-indexing method for substring matching. We adopted it (referred to as \underline{\tsf{Plain-SW}}) to process all the data trajectories.
Since \tsf{DITA} and \tsf{ERP-index} enumerate all subtrajectories, the whole datasets are impossible to index due to exceeding the main memory 
(e.g., \beijing~dataset generated 1.4~billion subtrajectories). 
We used a fraction of the dataset when these two methods were included in the competitors. %

We used the six WED instances introduced in \S~\ref{sec:prelim:ged} as similarity functions.
Instead of specifying a similarity threshold $\tau$ directly, we used a threshold ratio $\tau_{\tsf{ratio}} \in [0, 1]$.
Given a query $Q$ and $\tau_{\tsf{ratio}}$, we set $\tau:=\tau_{\tsf{ratio}}\sum_{q\in Q}c(q)$.
We used $\tau_{\tsf{ratio}}=0.1$ as the default value. 
For \surs, \lors, and \lcrs, costs are given by road lengths. The cost 
functions of the other WED instances are given in \S~\ref{sec:prelim:ged}. 
For \edr, we used $\varepsilon=0.001$. %
For \netedr, we set $\varepsilon$ to the median distance of edges in $E$.
For \neterp~deletion cost $G^{\tsf{(del)}}_\neterp$, we used $2M$. 
\revise{
For $\eta$ in Eq.~\eqref{eq:sub-nei}, we used $\eta=0$ for \lev, \edr, \surs, and \netedr, 
$10^{-4}$ multiplied by the median distance of a node and its nearest neighbor for \erp, 
and the median road length for \neterp 
\confversion{(see Appendix D of the extended version~\cite{extended-version} for the choice of $\eta$).}
\fullversion{(see Appendix~\ref{appendix:eta} for the choice of $\eta$).}
}

All evaluations were conducted on a workstation with Intel Core i9-7900X CPU (3.30GHz) and 64GB RAM.
All methods were implemented in \texttt{C++} (\texttt{g++} v.7.3.0) with the \texttt{-O3} option. We do not use multi-threading (for distributed baselines, we implemented centralized versions to compare algorithms themselves). All the algorithms were implemented in a main memory fashion.

\subsection{Effectiveness}
\label{sec:experiments:travel-time}
\subsubsection{Travel Time Estimation}
\begin{figure}[t]
    \centering
    \includegraphics[width=\linewidth,trim=1.5em 2.5em 2.5em 0]{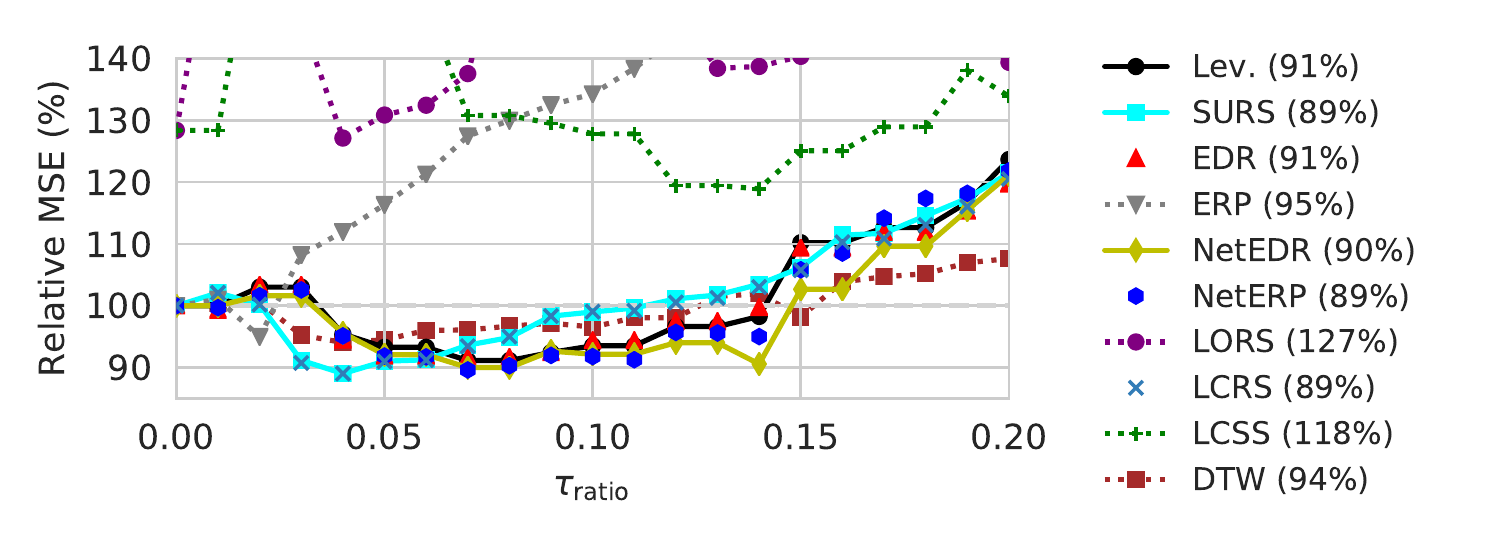}  %
    \caption{RMSE of travel time estimation (\beijing), best values reported on the right side.} 
    \label{fig:travel_time_functions}
\end{figure}

To demonstrate the effectiveness of WED and subtrajectory similarity search,
we first consider an on-the-fly travel time estimation task following the approach in \cite{EDBT19-Waury}. 
We use the \beijing dataset and sample 130 queries of length 60, where the numbers of exact match are 
less than 10 (i.e., the sparse case). 
The travel times corresponding to the subtrajectories that exactly match the query $Q$ are used as 
ground truth data. 
For estimation, we find the subtrajectories $\{P^{(id)}_{i:j}\}$ in the database similar to $Q$ under 
a threshold $\tau_{\text{ratio}}$. 
Since a trajectory $P^{(id)}$ may have multiple subtrajectories similar to $Q$, we pick the most 
similar one and break tie by the shortest one.
Then we compute the average of the travel time, $T^{(id)}_j\!\!-T^{(id)}_i$, over those similar to $Q$ 
as the estimated value. 
In order to evaluate the superiority of similarity search over exact match, we measure mean squared errors 
(MSEs) and report the relative value ($RMSE:= MSE(\tau_{\text{ratio}})/MSE(\text{exact})$). Since the ground 
truths are contained in both results of similarity search and exact match, we employ a leave-one-out 
\confversion{cross-validation}
\fullversion{cross-validation~\footnote{See Appendix~\ref{appendix:effectivenss} for details.}} by excluding one ground 
truth from the result set at a time. RMSE $< 100\%$ means similarity search is better than exact match.  
We compare the six WED instances in \S~\ref{sec:prelim:ged} with \dtw, \lors, \lcrs, and \lcss. 
Since \lors, \lcrs, and \lcss are defined on shared road segments, we convert them to equivalent 
distance functions. \dtw, \lors and \lcss are normalized to $[0, 1]$ such that  
$\tsf{DTW}(P,Q)\le\tau_\text{ratio}\sum_{i=1}^{|Q|-1}d(Q_i, Q_{i+1})^2$ and 
$\tsf{LORS}(P,Q)\ge (1-\tau_\text{ratio})\cdot \sum_{i=1}^{\size{Q}}w(Q_i)$ (the same for \lcss). 
Figure~\ref{fig:travel_time_functions} shows how the RMSE changes 
over $\tau_{\text{ratio}}$ (\lors and \lcss have RMSE $> 100\%$ at $\tau_{\text{ratio}} = 0$ because 
they do not count mismatching road segments in $P^{(id)}$). 
For most WED instances, similarity search performs better than exact match 
when $\tau_{\text{ratio}} \in [0.04, 0.14]$, showcasing the superiority of similarity search for 
travel time estimation on sparse data. To compare similarity functions, all the WED instances 
except \erp perform well, while \lors and \lcss deliver the worst performance. \surs achieves the 
smallest RMSE (89\%) among all the similarity functions. \netedr and \neterp are competitive when 
$\tau_{\text{ratio}}$ is large. These observations suggest that WED is useful for travel time 
estimation and these new WED instances are better than existing ones (\lev, \edr, and \erp). 
We also observe that \lcrs and \surs behave similarly because of similar semantics. 
An advantage of \surs over \lcrs is that efficient subtrajectory search under \lcrs has not been 
established and thus we enumerate all subtrajectories ($O(\sum_{k=1}^{N}\size{P}^{2}\size{Q})$-time), 
while \surs belongs to WED and can be efficiently processed. 
Next, we compare similar subtrajectory matching with whole matching. Since whole matching 
finds no result for most thresholds, we consider a top-$k$ setting for fair comparison. The 
RMSE is reported in Table~\ref{tab:travel_time_sub_whole} for \surs, the best function in 
the previous experiment. The result shows that the RMSE of subtrajectory matching is about half of 
the RMSE of whole matching, and the gap is more significant for small $k$.

\begin{table}[t]
    \centering
    \caption{RMSE of travel time (\surs, \beijing).}
    \small
    \begin{tabular}{r|ccccc} \hline
     $k$   & $5$  &  $10$  &  $15$  &  $20$  &  $25$  \\ \hline
    Subtrajectory & 92~\%  &  91~\%  &  102~\%  &  108~\%  &  116~\%  \\
    Whole   & 233~\%  &  221~\%  &  219~\%  &  220~\%  &  220~\%  \\ \hline
    \end{tabular}
    \label{tab:travel_time_sub_whole}
\end{table}

\subsubsection{Alternative Route Suggestion}
\revise{
Next we show the effectiveness through an alternative route suggestion task. Suppose a driver 
is planning to travel from an origin $u$ to a destination $v$ through a route $Q$, and the 
driver wants to find if there are variations of $Q$ as alternative routes. We can do this by 
retrieving subtrajectories from $u$ to $v$ similar to $Q$ from the database. To measure the 
preference of a route, we employ the route naturalness described in \cite{CST} \S7: drivers 
prefer routes that go directly towards the destination, and the log-likelihood of a route is 
proportional to the number of hops that get closer (in terms of road network distance) to the 
destination than ever. Following this idea, given a route $P$ such that $P_1 = u$ and 
$P_{\size{P}} = v$, we define its naturalness as the ratio of hops that get closer to $v$ than 
ever, i.e., $\frac{\size{C}}{\size{P} - 1}$, where $C = \set{(P_{i - 1}, P_i) \mid \min_{1 \leq 
j < i} d(P_j, v) > d(P_i, v)}$. If a route includes many inefficient detours, then the 
naturalness is low.}

\begin{figure}[t]
    \centering
    \includegraphics[width=\linewidth,trim=0 3em 0 0]{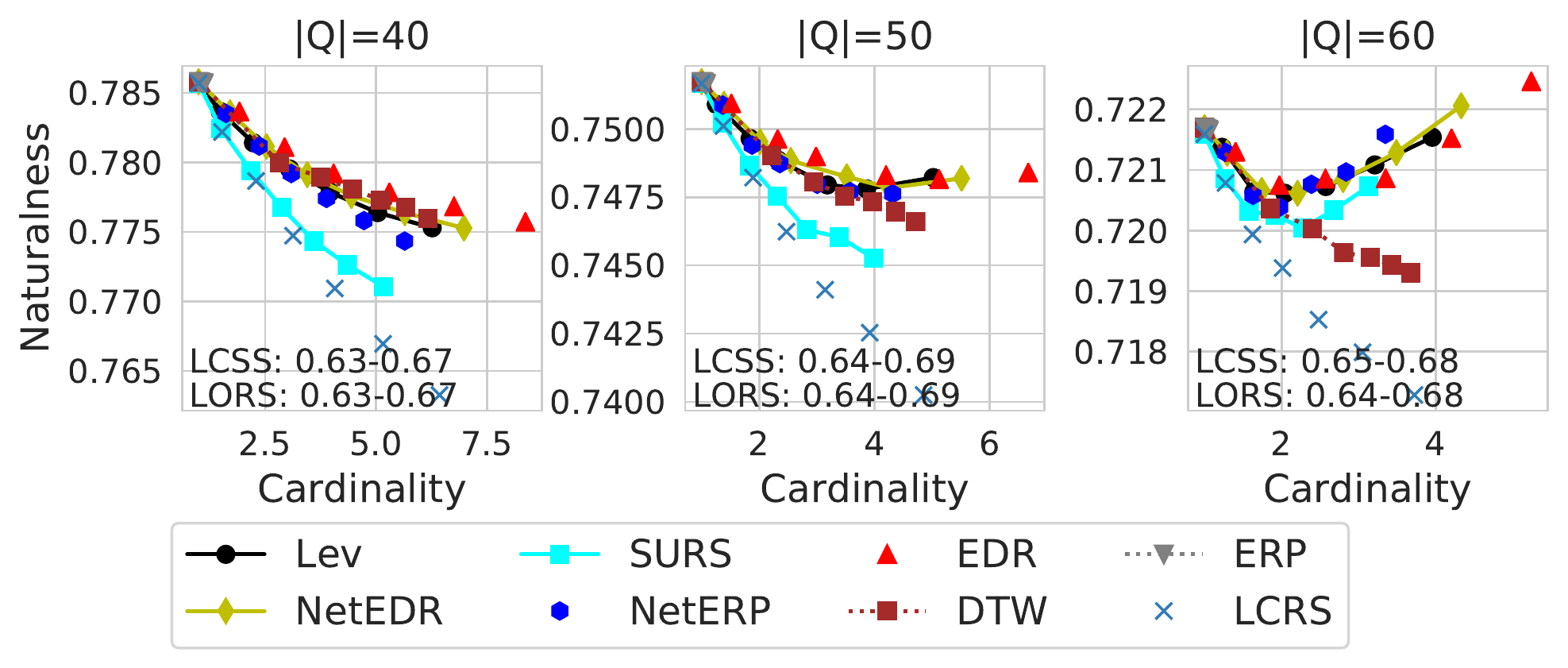}  %
    \caption{Naturalness of alternative routes suggested by similarity search $\tau_{\text{ratio}}\in[0,0.3]$ (\beijing).} 
    \label{fig:naturalness}
\end{figure}

\revise{
Figure~\ref{fig:naturalness} shows the naturalness of the routes suggested by subtrajectory 
similarity search under various similarity functions. The results are averaged over 3,000 
queries uniformly sampled from the \beijing dataset. We vary $\tau_{\text{ratio}}$ from 0 to 
0.3 at 0.05 interval and plot the cardinality (i.e., the number of suggested routes) and the 
naturalness. The cardinality increases w.r.t. $\tau_{\text{ratio}}$, but the rate depends on 
the similarity function. So we do not show $\tau_{\text{ratio}}$ explicitly. Among the six 
WED instances, \lev, \edr, \netedr, and \neterp deliver routes with high naturalness. \lcss, 
\lors, and \lcrs exhibit low naturalness because they measure common road segments but do not 
penalize inefficient detours. \dtw's naturalness is high for short queries but drops rapidly 
for long queries. Another interesting observation is that when query length is 50 or 60, the 
naturalness using WED instances, except \erp, first decreases w.r.t. the cardinality and then 
rebounds. This is because some highly natural routes involve shortcuts that are spatially 
distant from the queries. They cause large \dtw or \erp and hence are not identified using the 
two functions, while the WED instances with non-spatial distance as costs are capable of 
capturing these routes.}

\begin{figure*}[t]
    \centering
    \confversion{\includegraphics[width=.95\linewidth,trim=0 0.5em 0 0]{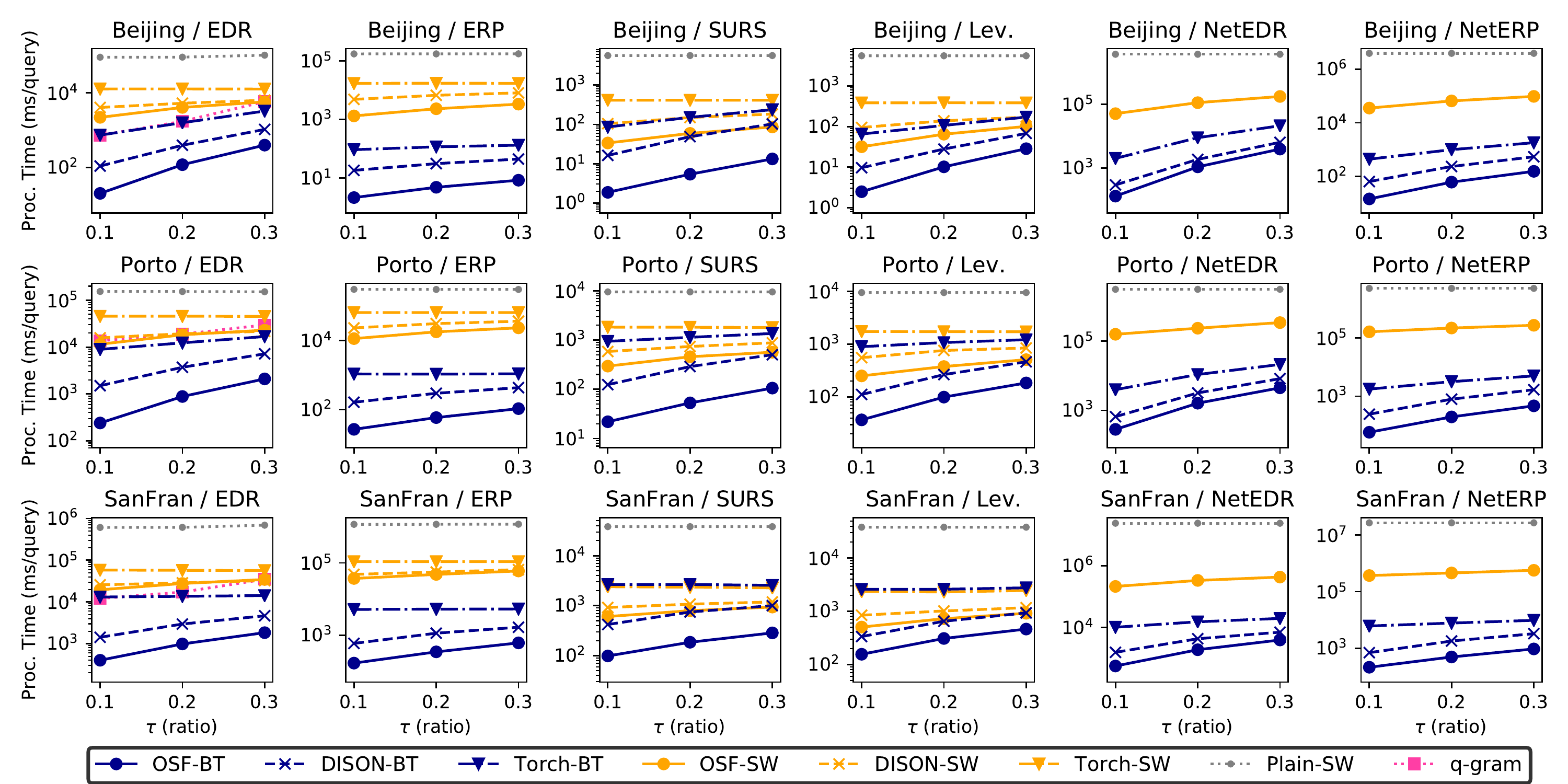}}
    \fullversion{\includegraphics[width=.95\linewidth,trim=0 2.5em 0 0]{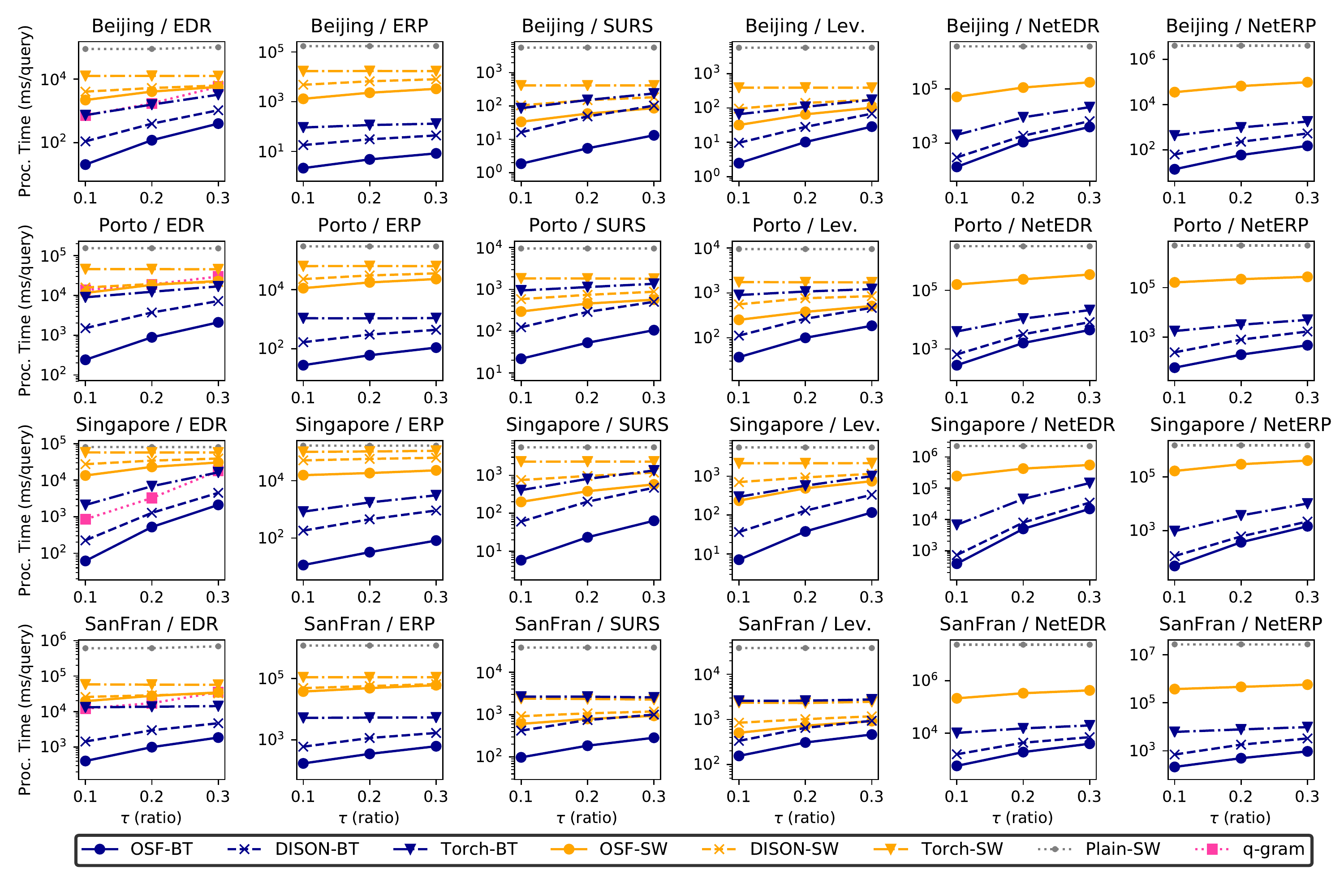}}
    \caption{Varying $\tau_{\tsf{ratio}}$: \tsf{OSF-BT} is our method (legend is shown at the bottom).} %
    \label{fig:varying_tau}
\end{figure*}

\subsection{Query Processing Time}
Following past related studies~\cite{DBLP:conf/kdd/WangZX14,EDBT19-Waury,SIGSPATIAL14-SPQ,TSAS18-SNT}, 
we randomly sampled subtrajectories from each dataset as queries. 
We set the default query length $\size{Q}$ to 60, whose path distance in real world 
ranges from 500~m to 40~km, with an average of 6.5~km, in line with \cite{DBLP:conf/kdd/WangZX14,EDBT19-Waury,yang2018urban}. 
Evaluation metrics were averaged over 100 queries, except for \tsf{Plain-SW}, whose 
processing time was averaged over 10 queries due to the computational cost.

\begin{figure*}[t]
  \begin{minipage}{0.495\linewidth}
    \centering
    \confversion{\includegraphics[trim=0 6em 0 0, width=\linewidth]{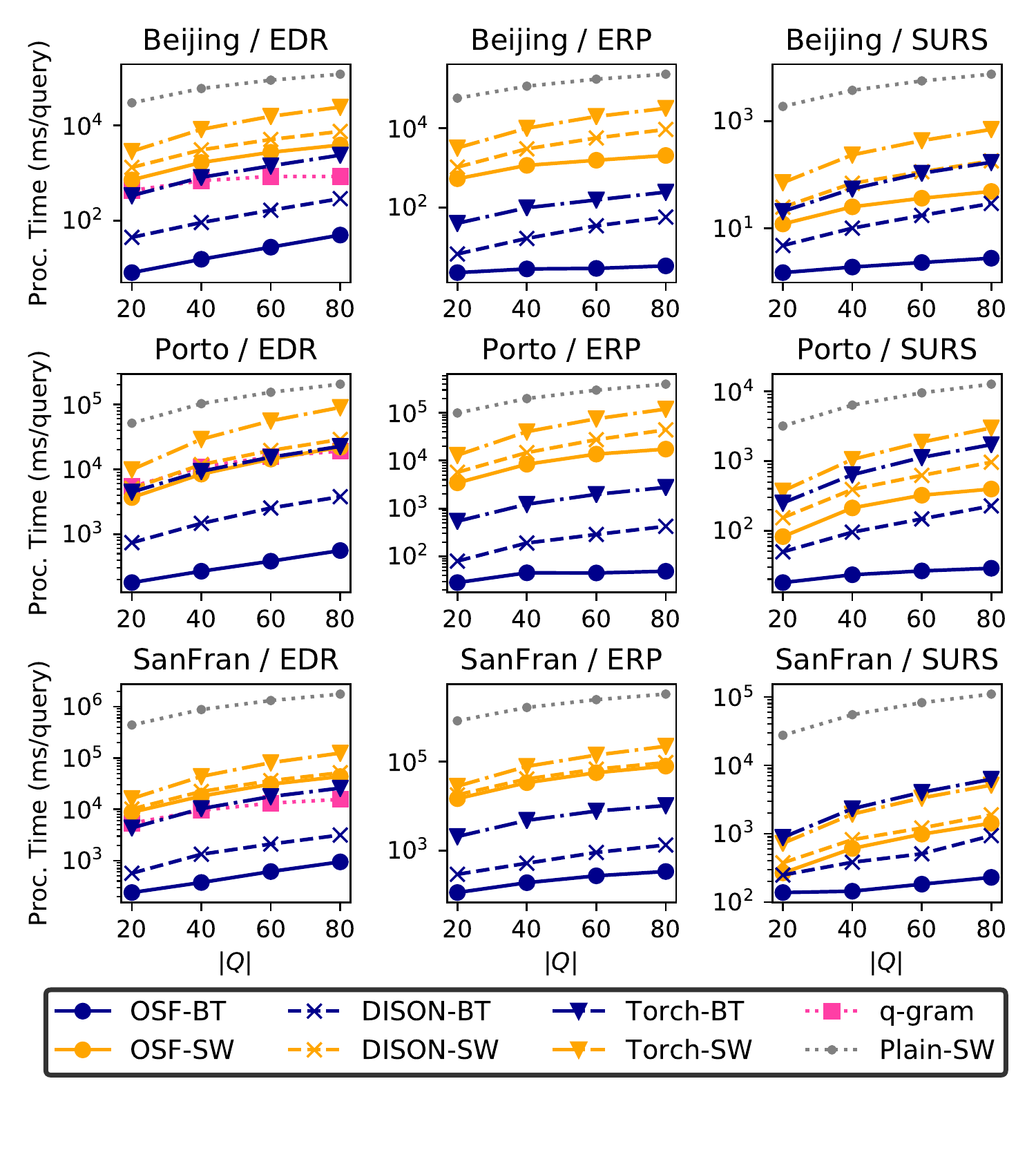}}
    \fullversion{\includegraphics[trim=0 8em 0 0, width=\linewidth]{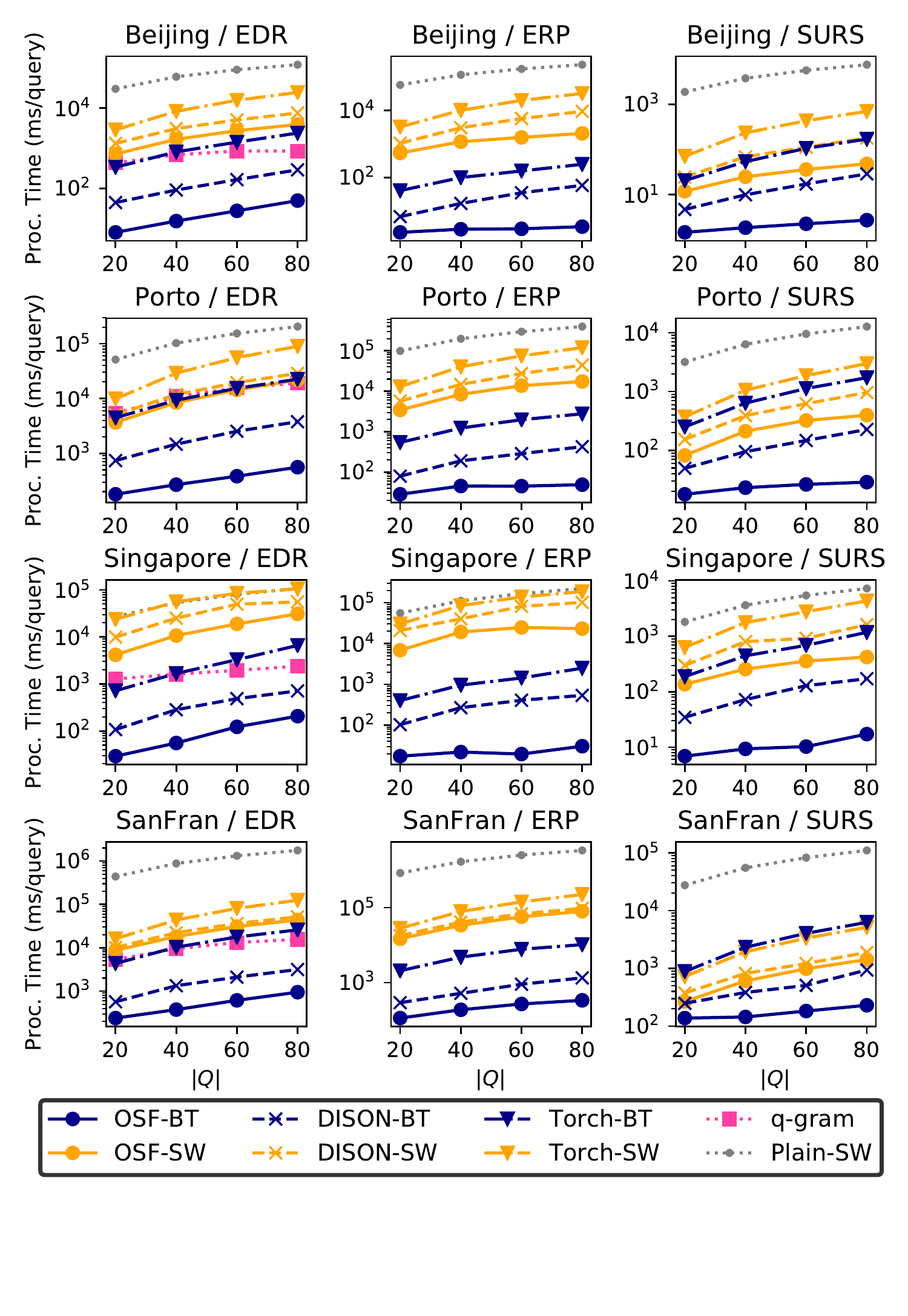}}
    \caption{Varying $|Q|$.}
    \label{fig:varying_Qlen}
  \end{minipage}
  \begin{minipage}{0.495\linewidth}
    \confversion{\includegraphics[trim=0 6em 0 0, width=\linewidth]{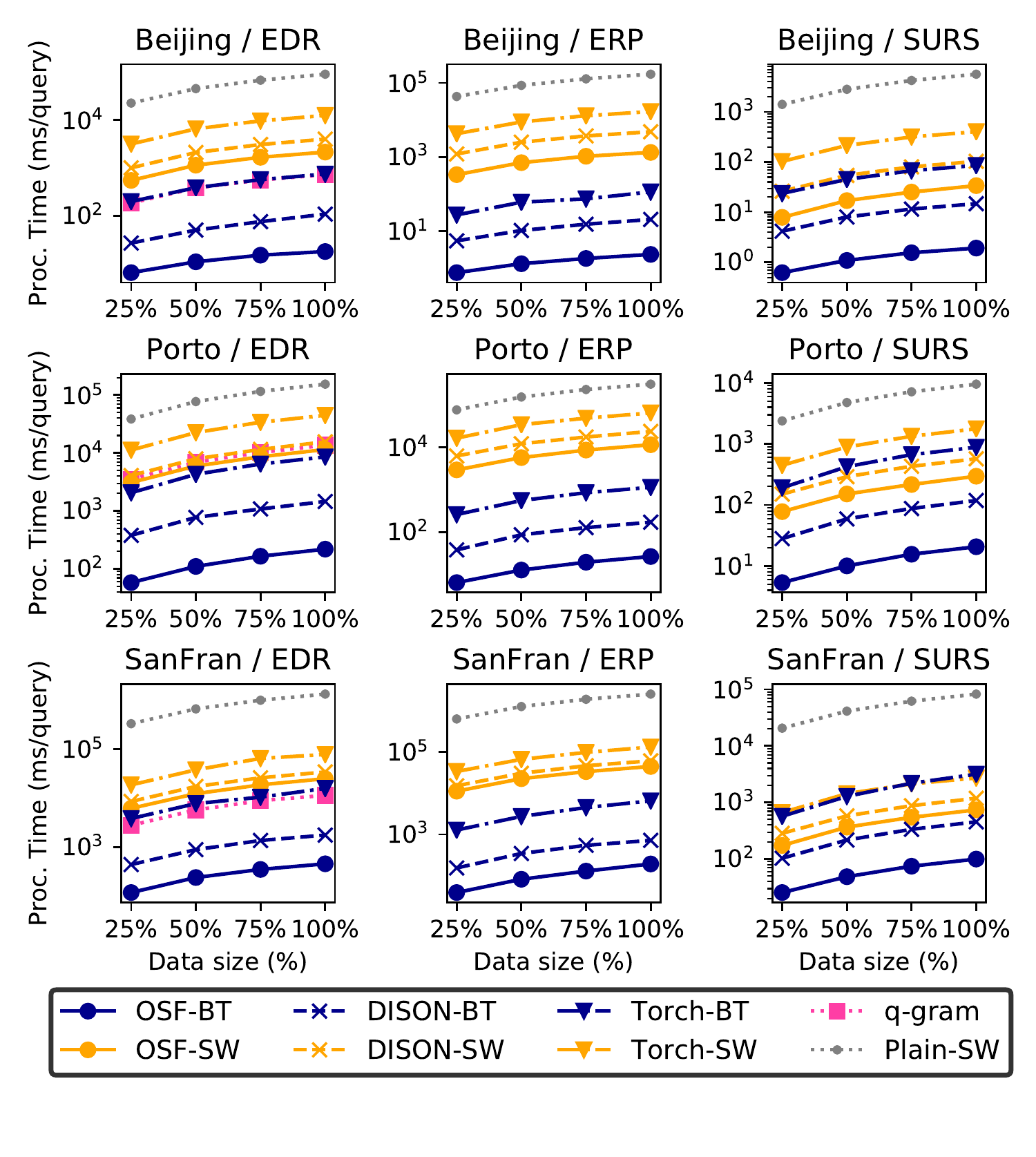}}
    \fullversion{\includegraphics[trim=0 8em 0 0, width=\linewidth]{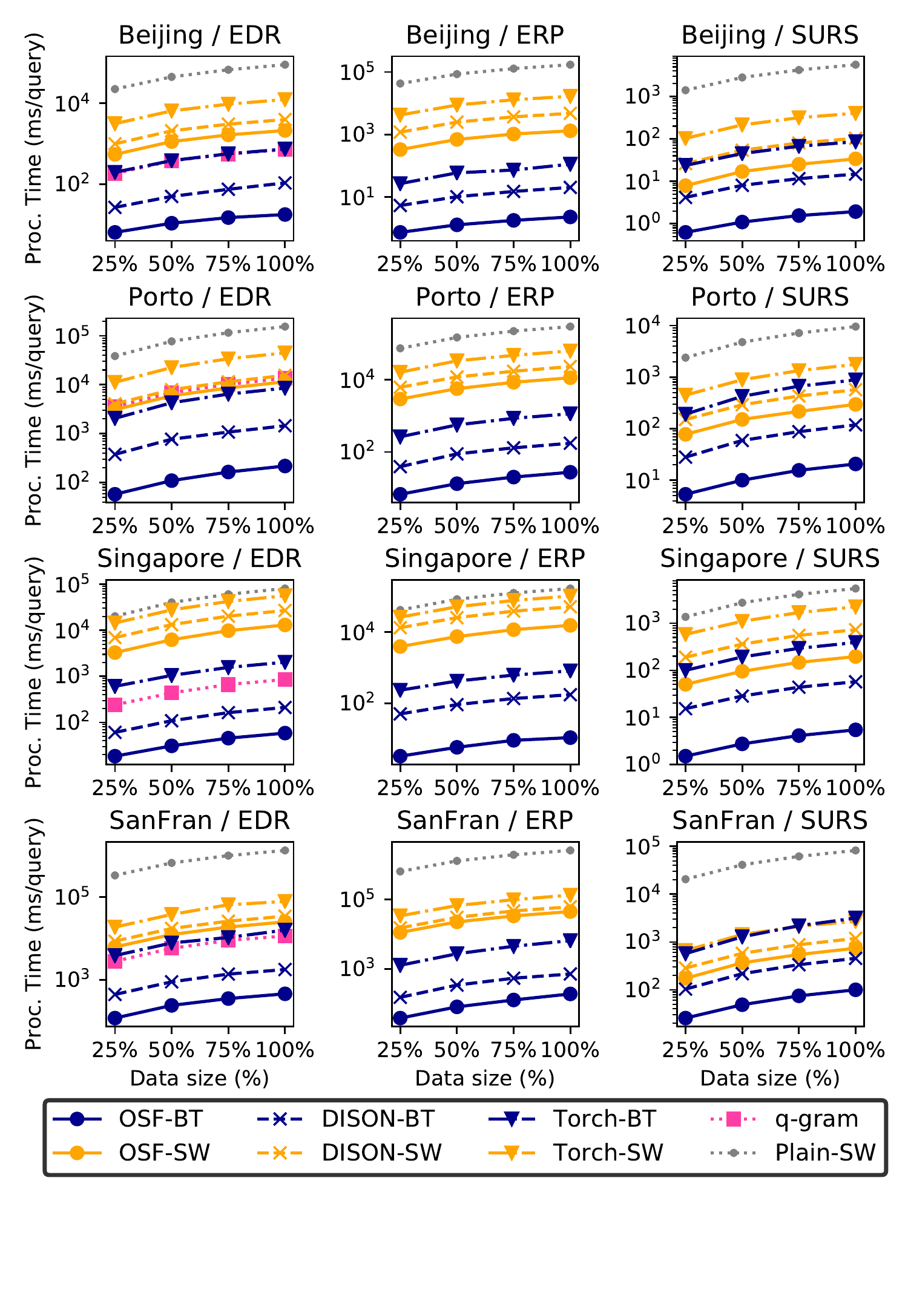}}
    \caption{Varying $\mathcal{T}$.}
    \label{fig:exp:ERP-index-varying-scalability}
  \end{minipage}
\end{figure*}

\begin{table}[t]
    \centering
    \caption{Running time breakdown (ms).}
    \begin{threeparttable}
    \small
    \begin{tabular}{c|r|rr|rr} \hline
     & \vspace{-0.29em}        & \multicolumn{2}{c}{Varying $\tau_\tsf{ratio}$} & \multicolumn{2}{|c}{Varying $\size{Q}$} \\
     & \raisebox{0.6em}{Default$^\dagger$} & 0.2 & 0.3 & 20 & 40 \\ \hline
     \textsc{MinCand} & 0.002  & 0.005   & 0.007   & 0.0005   & 0.001\\ 
     Index lookup & 0.070 & 0.259    & 0.443  & 0.039 & 0.055 \\ 
     Verify & 19.9 & 113.1 & 390.0 & 6.2  & 11.1 \\ \hline            
    \end{tabular}
     \label{tab:breakdown}
     \begin{tablenotes}
     \item[$\dagger$] {\scriptsize Default: $\tau_\text{ratio}=0.1$, $|Q|=60$, dataset size $=100\%$.}
     \end{tablenotes}
     \end{threeparttable}
\end{table}

We first investigate the effect of the similarity threshold $\tau_{\text{ratio}}$.
Figure~\ref{fig:varying_tau} shows that the proposed method \tsf{\GSF-BT} outperforms the other competitors. 
It responses in less than 2 seconds except for \netedr, and typically hundreds of milliseconds when $\tau_{\text{ratio}} = 0.3$.
Compared to \tsf{DISON-BT} and \tsf{Torch-BT}, which employ different filtering principles, our 
\tsf{\GSF-BT} always performs better and is up to 9 times faster than \tsf{DISON-BT} and 73 times 
faster than \tsf{Torch-BT}. 
Comparing our BT verification with SW, BT significantly improves the efficiency.
The impact of BT is more significant for \netedr~and \neterp, which involve relatively expensive computation 
in verification. For this reason, we omitted the results of \tsf{DISON-SW} and \tsf{Torch-SW} for \netedr~and \neterp~from Figure~\ref{fig:varying_tau}, which take at least 24~hours for computation for 100 queries.
These results indicate that both \tsf{\GSF} and \tsf{BT} improve the performance, and 
improvements are consistently observed whatever similarity function is used. 

\begin{figure*}
    \centering
    \includegraphics[width=0.487\linewidth, trim=0 2em 0 0]{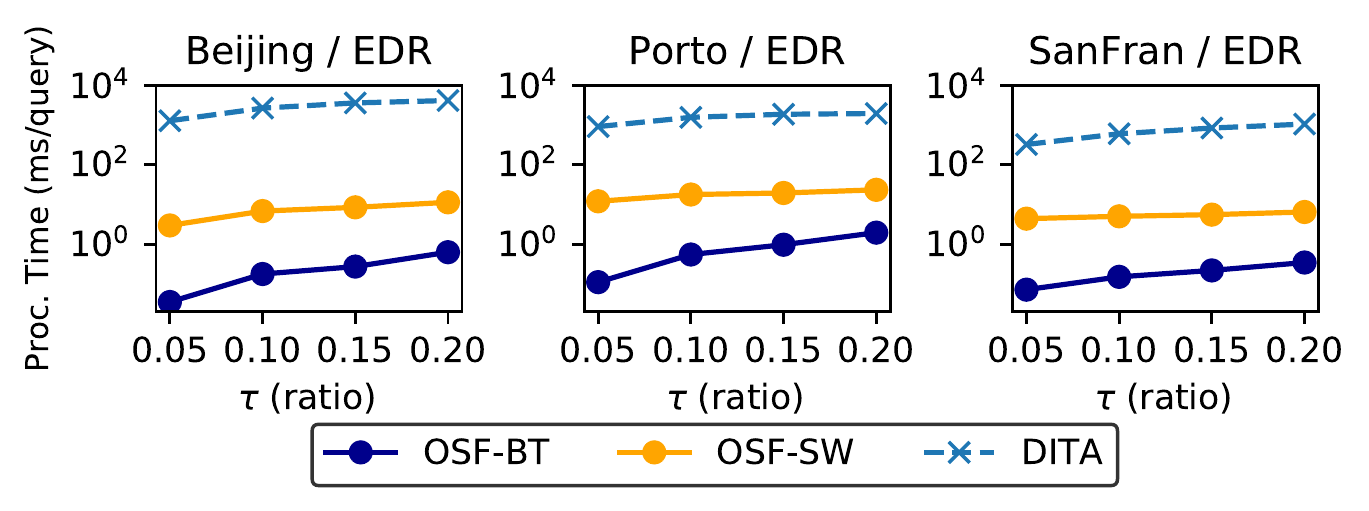}
    \includegraphics[width=0.49\linewidth, trim=0 2em 0 0]{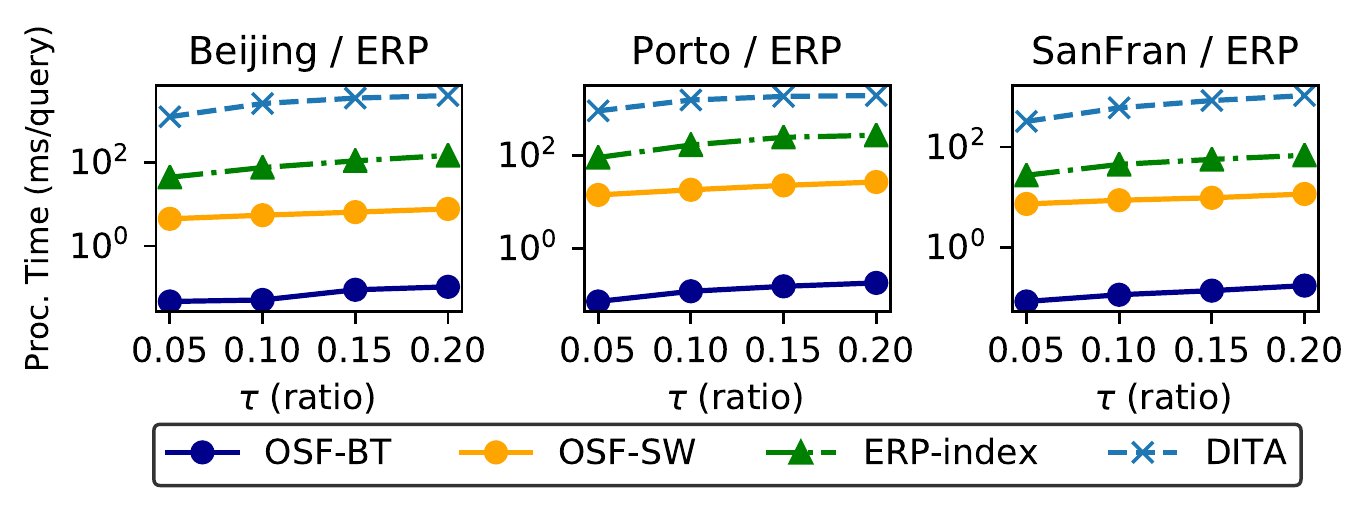}
    \caption{Comparison with baselines involving subtrajectory enumeration (Varying $\tau_{\text{ratio}}$; $\size{\database}=5000$; \edr/\erp).}
    \label{fig:exp:ERP-index-varying-tau}
    \vspace{1em}
    \includegraphics[width=0.487\linewidth, trim=0 2em 0 0]{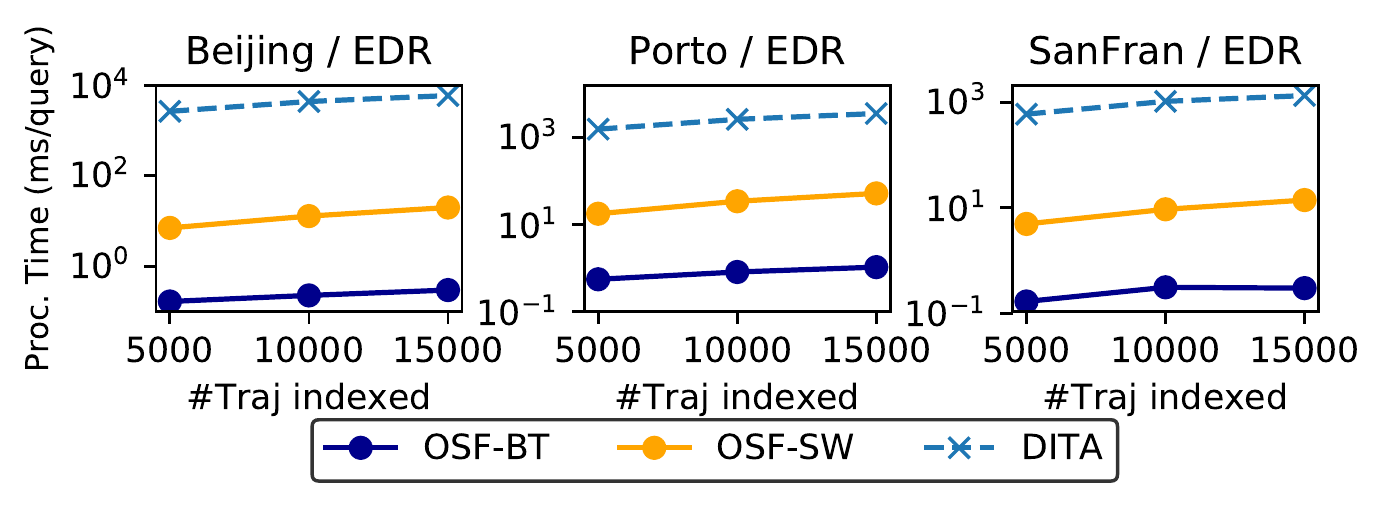}
    \includegraphics[width=0.49\linewidth, trim=0 2em 0 0]{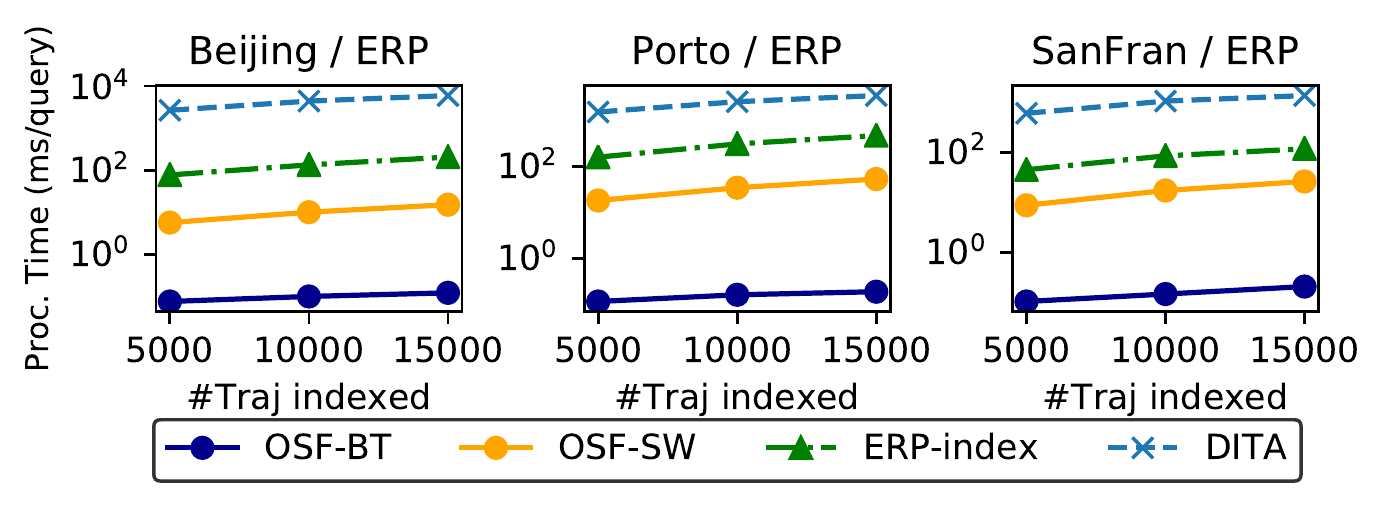}
    \caption{Comparison with baselines involving subtrajectory enumeration (Varying $\size{\database}$; $\tau_{\text{ratio}}=0.1$; \edr/\erp).}
    \label{fig:exp:ERP-index-varying-trajectory}
\end{figure*}

We vary the length of query $|Q|$ with $\tau_{\text{ratio}}=0.1$.
As shown in Figure~\ref{fig:varying_Qlen}, 
\tsf{\GSF-BT} is always faster than the others.
For larger $|Q|$, the processing time of \tsf{\GSF-BT} increases as well as the other methods.
The reason is two-fold:   
\begin{inparaenum} [(1)]
  \item the verification cost is proportional to $|Q|$; 
  \item in our setting, the similarity threshold $\tau$ increases as $|Q|$ increases, making the candidate set larger.
\end{inparaenum}

In Table~\ref{tab:breakdown}, we decompose the query processing time of \tsf{\GSF-BT} (on \beijing, \edr) to: 
\textsc{MinCand} computation, index lookup, and verification. 
Most time (around 99\%) is spent on verification, whose time increases with $\tau_{\text{ratio}}$ and $\size{Q}$. 
This is expected, because for each candidate we need only one index lookup but run a quadratic-time DP to verify it. 
\textsc{MinCand} computation is almost negligible since it runs in $O(|Q|^2)$ time, which does not depend on the dataset size. 

Figure~\ref{fig:exp:ERP-index-varying-scalability} shows query processing time when we vary the dataset size. 
All the methods scales linearly and \tsf{\GSF-BT} is consistently the fastest. 
In addition, we show results for \tsf{DITA} and \tsf{ERP-index}, which requires subtrajectory enumeration, 
on a fraction of the datasets where they can fit into the main memory; 
e.g., in order to store the randomly chosen 5,000 trajectories, the number of subtrajectories to be indexed is 
48M (\beijing), 37M (\porto), and 26M (\sanfran). 
The query processing time for this small dataset is shown in Figure~\ref{fig:exp:ERP-index-varying-tau} (varying $\tau_{\text{ratio}}$) 
and Figure~\ref{fig:exp:ERP-index-varying-trajectory} (varying dataset size).
Our method outperforms \tsf{DITA} and \tsf{ERP-index} by two orders of magnitude. 
This result indicates that applying whole matching methods to subtrajectory matching by enumerating all 
subtrajectories is impractical for large datasets.

\subsection{Filtering and Verification}
\label{sec:experiments:filtering-verification}
\begin{figure}[t]
    \centering
    \includegraphics[trim=0 2.5em 0 0,width=\linewidth]{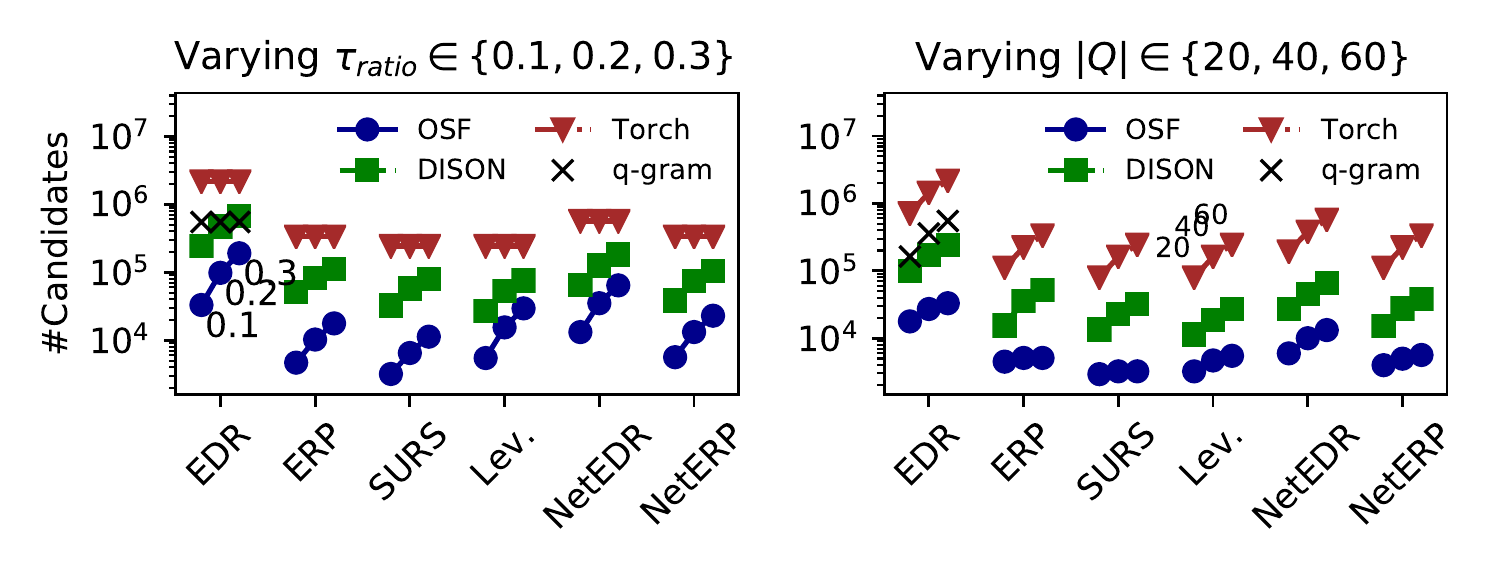}
    \caption{Number of candidate positions (\beijing).}
    \label{fig:22-candidate-varying-tau}
\end{figure}

\revise{We compare the filtering power by evaluating the candidate size $|\mathcal{C}|$. 
Note for all the competitors the candidates are in the form of $(id, j, i_q)$ for fair comparison. 
Figure~\ref{fig:22-candidate-varying-tau} shows: 
\begin{inparaenum} [(1)]
  \item \tsf{\GSF} consistently results in the best filtering power; its candidate 
  size is on average 3.4, 2.9, and 25 times smaller than \tsf{DISON}, \tsf{$q$-gram}, 
  and \tsf{Torch}, respectively. 
  \item \tsf{\GSF} shows good scalability against $|Q|$, because for long $|Q|$, the 
  item set in \textsc{MinCand} becomes large and thus the probability of including 
  ``good-value-for-the-price'' items becomes large.
\end{inparaenum}
We also compare with \tsf{DITA} and \tsf{ERP-index} on a fraction of the datasets. 
Their candidate sizes are on average 105 and 14 times \tsf{\GSF}'s, respectively.}

To separate the effect of local verification (\S~\ref{sec:bidirectional-search}) 
and the effect of caching with BT (\S~\ref{sec:verify:trie}), we evaluate
(i) unpruned position rate (UPR), and
(ii) cache miss rate (CMR),
which are respectively defined as
(i) the rate of DP columns that pass the early termination (\S~\ref{sec:bidirectional-search})
compared to SW, and %
(ii) the rate of DP columns where the \tsf{StepDP} procedure is actually called among the DP columns that pass the early termination. 
Table~\ref{tab:verify} shows results for the \beijing~dataset under \edr.
We observe both local verification and BT contribute to pruning. 
The rates increase with $\tau_{\tsf{ratio}}$ and $\size{Q}$ due to looser 
similarity constraint and longer query to verify, but decrease with dataset size 
due to more shared prefixes/suffixes. 
The total unpruned rate (TUR), defined by $\text{UPR}\times\text{CMR}$, shows small values, 
indicating the number of \tsf{StepDP} calls is far less than that by SW.

\begin{table}[t]
    \centering
    \caption{Evaluation of verification (\%).}%
    \begin{threeparttable}
    \small
    \begin{tabular}{c|r|rr|rr|rr} \hline
     & \vspace{-0.29em} & \multicolumn{2}{c}{Varying $\tau_\tsf{ratio}$} & \multicolumn{2}{|c}{Varying $\size{Q}$} & \multicolumn{2}{|c}{Varying $\size{\database}$} \\
     & \raisebox{0.6em}{Default$^\dagger$} & 0.2 & 0.3 & 20 & 40 & 25\% & 50\% \\ \hline
    UPR &  21.89 & 52.05 & 94.28 & 6.57 & 14.45 & 23.15 & 22.65\\ 
    CMR &  2.19 & 4.72 & 7.50 & 0.35 & 1.13 & 4.43 & 2.99\\ 
    TUR &  0.48 & 2.46 & 7.07 & 0.02 & 0.16 & 1.02 & 0.68\\ \hline            \end{tabular}
     \label{tab:verify}
     \begin{tablenotes}
     \item[$\dagger$] {\scriptsize Default: $\tau_\text{ratio}=0.1$, $|Q|=60$, dataset size $=100\%$.}    
     \end{tablenotes}
     \end{threeparttable}
     \vspace{-1.5em}
\end{table}

\subsection{Index Construction Time / Index Size}
Table~\ref{tab:index-stats} the index construction time and index size.
The index construction time of our postings lists is relatively fast.
For reference, we show results for methods involving subtrajectory enumeration (\tsf{ERP-index} and \tsf{DITA}).
Although these are results with only 5,000 trajectories, the index construction time and index size are larger than ours except on the \sanfran dataset.

\begin{table}[t]
    \centering
    \caption{Index construction time / index size.
    }
    \label{tab:index-stats}
    \begin{threeparttable}
    \small
    \begin{tabular}{l|p{1.8cm}|p{1.8cm}|p{1.8cm}}\hline
                                      & \beijing & \porto & \sanfran \\ \hline
        \tsf{OSF-BT}$^\dagger$                & 7s / 0.59~\textsc{gb}   & 10s / 1.02~\textsc{gb} &    79s / 8.63~\textsc{gb}\\
        \tsf{$q$-gram}                      & 15s / 0.59~\textsc{gb}  & 19s / 1.01~\textsc{gb} &   269s / 8.55~\textsc{gb}\\ \hline\hline
        $\downarrow$ {\scriptsize Tiny dataset} & & & \\
        (\tsf{ERP-index}$^{\dagger\!\dagger}$)  & 23s / 2.6~\textsc{gb} & 17s / 1.9~\textsc{gb} & 14s / 1.4~\textsc{gb}  \\
        (\tsf{DITA}$^{\dagger\!\dagger}$)       & 60s / 1.79~\textsc{gb} & 36s / 0.71~\textsc{gb} & 31s / 0.15~\textsc{gb} \\ \hline
    \end{tabular}
    \begin{tablenotes}
    \item[$\dagger$] {\scriptsize \tsf{DISON} and \tsf{Torch} have the same time/size as \tsf{OSF-BT}.} 
    \item[$\dagger\!\dagger$] {\scriptsize Reference values with only 5,000 trajectories. \tsf{DITA} construction depends on the similarity function. Here we showed a result for \erp.} 
    \end{tablenotes}
    \end{threeparttable}
    \vspace{-1em}
\end{table}

\fullversion{
\subsection{Temporal Constraints}
We show the results with temporal constraints of type $[T_i,T_j]\cap I\ne\emptyset$, where $I$ is a 
query time interval. We vary temporal selectivity (e.g., 1\% temporal selectivity means 
$I=[ts_{\min}, ts_{1\%}]$, where $ts_{\min}$ is the minimum timestamp and 
$ts_{1\%}$ is the 1\% quantile). For pruning, temporal constraints are checked 
after candidate generation (referred to as \tsf{TF}, see \S~\ref{sec:index:extension}). 
With this pruning, the processing time scales almost linearly with temporal selectivity. 
We compare with the method that checks temporal constraints as postprocessing (\tsf{no-TF}).  Figure~\ref{fig:temporal_selectivity} shows that \tsf{TF} is faster by one order of 
magnitude. The gap is more significant when temporal selectivity is low.

\begin{figure}
    \centering
    \includegraphics[width=0.98\linewidth,trim=0 2.5em 0 3mm]{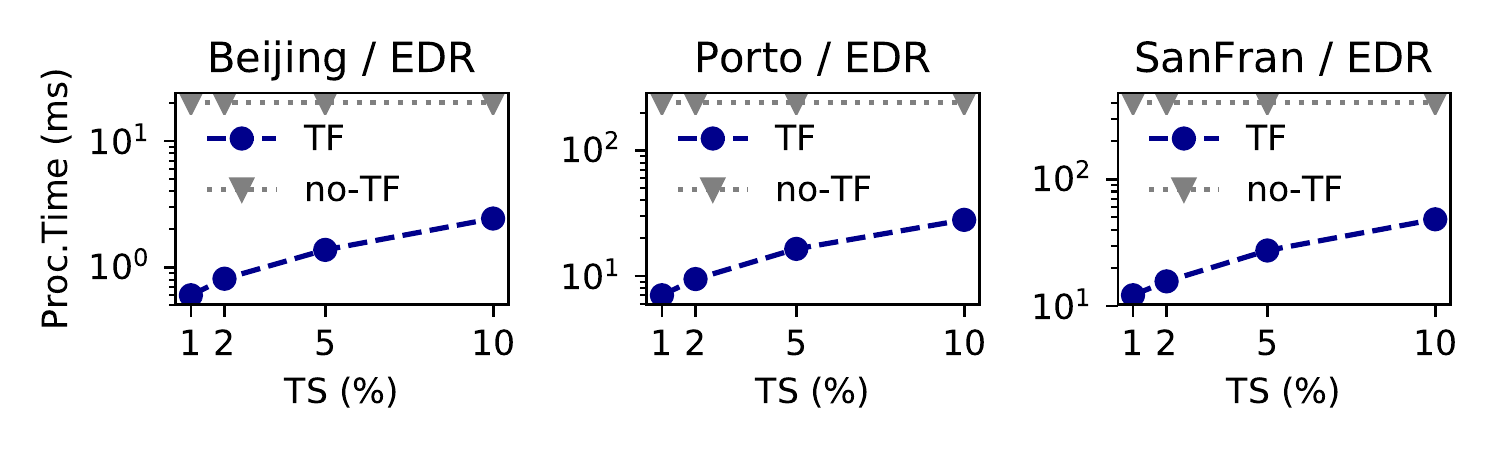}
    \caption{Varying temporal selectivity (TS).}%
    \label{fig:temporal_selectivity}
\end{figure}
}

\section{Related Work}
\label{sec:related-work}
\myparagraph{Trajectory Similarity Functions}
Trajectory similarity functions can be classified into: 
\begin{inparaenum} [(1)]
 \item coordinate-aware similarity functions are defined based on spatial coordinates of trajectories \cite{DBLP:conf/icde/YiJF98,Shim2003,EDR,ERP,FeiFei2017,CST} and 
 \item network-aware similarity functions employ network features, such as travel costs \cite{ShuoShang2017,VLDB14-Personalized,SIGIR18-Torch,Won2009,Xia2011,Evans2013,Hwang2006,Tiakas2009,ICDE18-DISON}.
\end{inparaenum}
Coordinate-aware similarity functions include dynamic time warping (\dtw), edit distance with real penalty (\erp) \cite{ERP}, edit distance on real sequence (\edr) \cite{EDR}, edit distance with projections (\edwp) \cite{EDwP}, and Fr\'{e}chet distance.
Their pros and cons were investigated in \cite{SIGIR18-Torch,EDR,ERP,ICDE18-DISON,VLDBJ-survey}.
\erp and \edr are WED instances. \dtw, \edwp, and Fr\'{e}chet distance are not. %
Recent effort aimed to learn deep trajectory representations~\cite{DBLP:conf/icde/LiZCJW18} or metrics~\cite{DBLP:conf/icde/YaoCZB19} to reduce the computation of similarity to linear time. 
For network-aware similarity functions, 
a natural way is to measure the shared or unshared edges. 
Weighted Jaccard distance~\cite{Xia2011} and weighted Dice distance~\cite{Won2009} 
are order-insensitive functions (i.e., edge ordering not incorporated). 
Longest common subsequences (\lcss)~\cite{DBLP:conf/icde/VlachosGK02,DBLP:conf/sigmod/MorseP07}, 
Longest overlapping road segments (\lors)~\cite{SIGIR18-Torch}, and longest common road segments (\lcrs)~\cite{ICDE18-DISON} are order-sensitive functions, 
while they do not belong to 
\confversion{WED.}
\fullversion{WED~\footnote{See Appendix~\ref{appendix:lors-lcrs-surs} for semantic comparison of \lors, 
\lcrs, and \surs.}.}
Another strategy is to incorporate shortest path distances between vertices~\cite{ShuoShang2017,VLDB14-Personalized,Evans2013,Hwang2006,Tiakas2009}.
\myparagraph{Trajectory Indexing}
Although much attention has been gathered to indexing for non-constrained trajectories (see surveys \cite{TrajSurveyPart1, TrajSurveyPart2}), indexing methods for trajectories in road networks are also studied actively \cite{MONTree, SIGSPATIAL14-SPQ, SPNET,  TSAS18-SNT, SIGIR18-Torch, ICDE18-DISON}.
Chen~et~al. proposed \erp~\cite{ERP} and \edr~\cite{EDR} along with the query processing algorithms. 
Wang~et~al.~\cite{SIGIR18-Torch} and Yuan and Li~\cite{ICDE18-DISON} proposed algorithms to support various similarity functions but they were designed for whole matching (though can be adapted for subtrajectories, see our experiments). 
Furthermore, their filtering policies are different from ours (based on scanning postings lists for all symbols~\cite{SIGIR18-Torch} or prefix symbols~\cite{ICDE18-DISON} of a query). 
Shang et al.~\cite{SIGMOD18-DITA} and Xie et al.~\cite{FeiFei2017} proposed distributed systems for similarity search under \dtw and discrete Fr\'{e}che/Hausdorff distances, respectively.
These methods \cite{SIGMOD18-DITA, FeiFei2017} support only whole matching. Pivot points were proposed in \cite{SIGMOD18-DITA} for pruning. The differences from our method are: 
\begin{inparaenum} [(1)]
  \item The pivots are selected by turning points or the distance to neighbor points/origin/destination, while our $\tau$-subsequence is chosen by a selectivity optimization algorithm. 
  \item In contrast to pivot points, our filtering is designed towards the general case of WED, and thus does not require individual adaptation for specific similarity functions.
\end{inparaenum} 
\myparagraph{String / Time-series Similarity Search}
Edit distance has been employed for string similarity search. 
Bouding techniques (e.g., by $q$-grams~\cite{DBLP:journals/tods/Qin0XLLW13,DBLP:conf/sigmod/DengLF14,DBLP:journals/tkde/WeiYL18}) 
are widely used. 
In bioinformatics, three types of similarity search methods %
are used: global \cite{NW}, local \cite{SW, BWA-SW, ALAE}, and semi-global 
alignments \cite{BWA}. The Smith-Waterman algorithm~\cite{SW} can be used for WED. 
Trajectories also belong to time series data, whose similarity is often measured by 
Euclidean distance, \dtw, \erp, or \lcss. 
Common indexing methods are based on lower bounding~\cite{Assent2008,Keogh2005}.
\fullversion{Substring similarity search can be carried out by indexing methods based on suffix arrays (or suffix-tries)~\cite{BWA-SW,CiNCT}.
Our BT method can be regarded as a method that dynamically builds prefix- and suffix-tries for verification instead of a static trie.}

\section{Conclusion and Future Work}
\label{sec:conclusion}
We tackled subtrajectory similarity search under WED, a class of similarity function 
that includes several important similarity functions. For efficient search, we 
proposed \filtername, which involves a discrete optimization problem to choose the 
optimal subsequence. Based on this technique, we developed an algorithm using 
filter-and-verify strategy. We designed a local verification method equipped with 
bidirectional tries. We showed the effectiveness of WED and the superiority of our 
solution over alternative methods. Interesting future directions include supporting 
more general class of similarity functions, developing a distributed indexing method, 
and more sophisticated treatment of temporal information.  
\myparagraph{Acknowledgments}
This work was supported by JSPS 16H01722, 17H06099, 18H04093, 19K11979, and NSFC 61702409.  
\balance 

\bibliographystyle{abbrv}
\bibliography{ref}

\fullversion{
\appendix
\label{appendix}
\section{Pseudo-codes}
\label{appendix:pseduo-codes}

Algorithm~\ref{algo:sw} describes the SW algorithm for finding the best substring $P_{s,t}\substr P$ that minimizes $\wed(Q,P_{s:t})$.
The minimum operation at Line~8 is taken based on the first elements, i.e., $(a,b,c)$.
Note that the matrix $K$ memorizes the first subscript $s$ of the current best substring; this technique was used in \cite{Sakurai2007}.

\begin{algorithm}
 \small
 \caption{$\textsf{Smith-Waterman}(Q, P, \wed)$}
 \label{algo:sw}
 \SetKwInOut{Input}{input}
 \SetKwInOut{Output}{output} \DontPrintSemicolon
 \Input{Query $Q\in\Sigma^*$; Trajectory $P\in\Sigma^*$}
 \Output{The best matching substring $P_{s:t}$ with $\wed(Q,P_{s:t})$}
 $D,K\leftarrow (|Q|+1)\times (|P|+1)$ zero matrices\;
 $D_{i,0}\leftarrow \sum_{i'=1}^i\del(P_{i'})$ \quad ($1\le i\le|Q|$)\;
 $\textit{BestSoFar}\leftarrow\infty$\;
 \For{$j$ \emph{\textbf{in}} $1..\size{P}$}{
     $K_{0,j}\leftarrow j$\;
     \For{$i$ \emph{\textbf{in}} $1..\size{Q}$}{
         $(a,b,c)\leftarrow(D_{i-1,j-1}+\sub(Q_i, P_j), D_{i-1,j}+\del(P_j), D_{i,j-1}+\ins(Q_i))$\;
         $(D_{i,j}, K_{i,j})\!\leftarrow\!\min\{(a, K_{i-1,j-1}), (b,K_{i-1,j}), 8c,K_{i,j-1})\}$\; %
         \If{$D_{i,j}<\textit{BestSoFar}$}{
              $\textit{BestSoFar}\leftarrow D_{i,j}$\;
              $(s,t)\leftarrow (K_{i,j},j)$
         }
     }%
 }
 \Return{\{$P_{s:t}$, \textit{BestSoFar}\}}
\end{algorithm}

\section{Proofs}
\label{appendix:proofs}

\myparagraph{Theorem~\ref{thm:gsf}}
\begin{proof}%
\revise{
Consider converting $Q$ into $P'\substr P$ with edit operations.
Focusing on $q\in Q$, if $P'\cap B(q)=\emptyset$, this $q$ must be deleted or substituted to another symbol $q'$ not included in $B(q)$.
Therefore, we must pay a cost at least 
$c(q)$, which is the 
smallest substitution cost from $q$ (note: deletion is considered as $\text{sub}(q,\varepsilon)$).
Similarly, if $P'\cap B(Q')=\emptyset$, we must pay a cost at 
least $c(Q'):=\sum_{q\in Q'}c(q)$.
So, if $P'\cap B(Q')=\emptyset$ and $c(Q')\ge\tau$, we cannot have $\wed(P',Q)<\tau$.
}
\end{proof}

\myparagraph{Proposition~\ref{theo:np-hard}}
\begin{proof}
\revise{
The minimum knapsack problem (MKP)~\cite{min-knapsack} 
is specified by a tuple $(K, \{W_k\}_{k=1}^K, \{V_k\}_{k=1}^K, D)$. $K$ is the number of items; 
$W_k$ is the weight of an item and $V_k$ is its value. The goal is to select a minimum weight subset of items 
$S \subseq [\![K]\!]$, such that the total value is no less than a demand $D$. Formally, 
\begin{align*}
    \min_{S \subseq [\![K]\!]} \sum_{k \in S} W_k,\quad 
    \text{subject to}\quad \sum_{k \in S} V_k \ge D. 
\end{align*}
}

\revise{
\textsc{MinCand} is specified by a tuple $(\Sigma, Q, \{n(b)\}_{b\in\Sigma}, \eta, \tau, \sub)$. $\Sigma$ is 
the alphabet; $Q$ is the query; $n(b)$ is the frequency of $b$; $\eta$ and $\tau$ are the thresholds of 
substitution neighbor and similarity search, respectively; $\sub$ specifies the WED cost functions. 
}

\revise{To prove the NP-hardness, it is sufficient to show that any MKP instance can be converted into a \textsc{MinCand} instance \emph{by constructing a specific instance of \textsc{MinCand}.}}
\revise{
Given any MKP instance $(K, \{W_k\}, \{V_k\}, D)$, we construct such a \textsc{MinCand} instance as follows.  
Let $\Sigma=\{1,2,\cdots,2K\}$; $Q=[1,2,\cdots,K]$; $n(k)=W_k$ if $k\in[\![K]\!]$, or $0$ otherwise; 
$\eta=\min_{k\in[\![K]\!]}V_k/2$; $\tau=D$. We define $\sub$ as 
\begin{align*}
    \sub(k,k')={\footnotesize
    \begin{cases}
    0, & k=k';\\
    V_{\min\{k,k'\}}, & |k-k'|=K;\\
    +\infty, & \text{otherwise}.
    \end{cases}
    }
\end{align*}
}

\revise{
We show that the decision versions of the two optimization problems return the same result (true/false).
By the above assignment, we have the substitution neighbor $B(k)=\{k\}$, $\forall k\in[\![K]\!]$. 
Hence, $c(k)=\min_{k'\in \Sigma\setminus\{k\}}\sub(k,k')=V_k$, $\forall k\in[\![K]\!]$.
Therefore, selecting a minimum weight subset of items for any MKP instance is exactly selecting a minimum 
frequency subsequence for a \textsc{MinCand}. 
}
\end{proof}

\myparagraph{Proposition~\ref{theo:approx-ratio}}
\begin{proof}
We show that any \textsc{MinCand} instance is a certain equivalent instance of MKP. 
First, we rewrite \textsc{MinCand} \eqref{eq:mincand-obj} by letting $K=|Q|$, $W_k=\sum_{q\in B(Q_k)}n(q)$, 
$V_k=c(Q_k)$, and $D=\tau$. Note that they are constant once a \textsc{MinCand} instance is fixed.
This leads to the following optimization problem:
\begin{align*}
    \min_{S \subseq [\![K]\!]} \sum_{k \in S} W_k,\quad 
    \text{subject to}\quad \sum_{k \in S} V_k \ge D, 
\end{align*}
which is of the same form as MKP.
Hence, we can use any (approximation) algorithm for MKP to solve \textsc{MinCand}.
Therefore, the 2-approximation property 
holds because Algorithm~\ref{algo:primaldual} is the 2-approximation algorithm for MKP~\cite{min-knapsack}.
\end{proof}

\myparagraph{Proposition~\ref{theo:global-optim}}
\begin{proof}
\revise{
We show that when $c(q) = c'$ is constant, Algorithm~\ref{algo:primaldual} chooses \topk elements with the smallest $N_q$ (i.e., 
frequency) values in $Q$, and thus the chosen $k$ elements are optimal. We prove this by induction. 
\begin{inparaenum} [(1)]
  \item In the first iteration, by the definition of $v_q$ (Line 4), the element with the smallest $N_q$ is chosen 
  (Line 5). 
  \item Assume that after the $i$-th iteration, the top-$i$ elements with the smallest $N_q$ values are 
  chosen. 
  \item We show that the $(i + 1)$-th element is chosen for the $(i + 1)$-th iteration: 
  In Line 4, we compute the $v_q$ value of each $q$. Because the denominator $\min\{c(q), \tau-c(Q')\} 
  = \min\{c', \tau-c(Q')\}$ is equal for every element $q \in Q\backslash Q'$, we choose the element with the 
  smallest $v_q \propto N_q - w_q$ in $Q\backslash Q'$ in Line 5. In Line 6, $w_q$ is incremented by $\min\{c(q), 
  \tau-c(Q')\} \cdot v_{q^*} = \min\{c', \tau-c(Q')\} \cdot v_{q^*}$ in each iteration. Because this increment is 
  equal for every element $q \in Q\backslash Q'$, we have an equal $w_q$ for all $q \in Q\backslash Q'$. Therefore, 
  in Line 5, we choose the element with the smallest $N_q$ in $Q\backslash Q'$, i.e., the $(i + 1)$-th element in 
  $Q$ ranked by $N_q$.
\end{inparaenum}
}
\end{proof}

\myparagraph{Lemma~\ref{theo:partition-dp}}
\begin{proof}
The statement means that, if $\wed(P_{s:t}^{(id)},Q)<\tau$, then there exists at least one candidate $(id, j, i_q)\in\mathcal{C}$ such that 
\begin{inparaenum} [(1)]
  \item $s\le j\le t$, and
  \item $Q_{i_q}\in Q'$ is \emph{substituted} (including a match) to $P_j^{(id)}\in B(Q_{i_q})$ in the optimal alignment ($Q'$ is the $\tau$-subsequence of $Q$). 
\end{inparaenum}
Suppose these two conditions are not true. Then every $Q_{i_q}\in Q'$ is deleted or substituted to a symbol not included in $B(Q_{i_q})$. By Theorem~\ref{thm:gsf}, 
$\nexists P_{s:t}^{(id)} \substr P$ such that $\wed(P_{s:t}^{(id)},Q)<\tau$, hence violating the first assumption that $\wed(P_{s:t}^{(id)},Q)<\tau$. 
\end{proof}

\section{Detailed Setup for Competitors}
\label{appendix:competitors}

\myparagraph{\tsf{DISON}}
It can be adapted to our problem by regarding its candidate generation process as a realization of 
$\tau$-subsequence $Q'$, though it might not be optimal in terms of candidate size. We derive a 
candidate condition by defining $Q'$ as the shortest prefix such that $\sum_{k=1}^i c(q_k)\ge\tau$, 
where $i$ is the prefix length. Candidates are generated by scanning postings lists~\footnote{To
guarantee the correctness, all the competitors in the experiments need to process not only $q\in Q$ 
but also substitution neighbors $B(q)$.} $L_q$ for a prefix $Q'$ of $Q$.

\myparagraph{\tsf{Torch}}
Candidates are generated by scanning postings lists $L_q$ for every $q\in Q$. 

\myparagraph{\tsf{DITA}}
For every subtrajectory $P$ of each data trajectory, pivots $P'\subseq P$ of length $K$ are chosen. 
The lower bound for \WEDs is $LB_{\text{pivot}}(P',Q):=\sum_{p\in P'}\min_{q\in Q\cup\{\varepsilon\}}\sub(p,q)\le \wed(P,Q)$.
Then, the pivots $P'$ are stored in a trie with the original trajectory ID.
For \edr, we chose frequent symbols as pivots to keep the trie compact. 
For \erp, we chose points with large deletion cost as pivots. 
These choices performed better than other options (e.g., random, small deletion cost, 
or the angle-based pivots in \cite{SIGMOD18-DITA}). 
For pivot size, we varied $K\in\{5, 10, 20\}$ and selected $K=10$, which 
resulted in the fastest processing.
\\[.5em]
\myparagraph{$q$-gram indexing for \edr}
We store each data trajectory as usual in $q$-gram inverted indexes (without substring enumeration).
For querying, we do as follows for each $q$-gram $x\in\Sigma^q$ in $Q$: 
We enumerate $x'\in\Sigma^q$ that matches $x$.
For each trajectory ID $id$, we count how many times $id$ appears in total 
in the inverted indexes over $x'$ using hash table $H[id]$.
Then, trajectories with $H[id]\ge |Q|-q+1-\tau q$ are verified by SW.
We use $|Q|$ as a lower bound of $\max\{|P|,|Q|\}$.

\section{Choice of $\eta$}
\label{appendix:eta}
\lev, \edr, and \netedr are unit-cost functions, and thus we set $\eta = 0$. Other values of $\eta$ are meaningless 
(either equivalent to $\eta = 0$ or making $B(q) = \Sigma$, $\forall q \in Q$). 

We also set $\eta = 0$ for \surs, which uses road lengths as costs. An $\eta > 0$ will cause $B(q)$ that contains road 
segments $q'$ spatially distant from $q$, because $\sub(q, q') = w(q) + w(q')$ are small for short road segments. This 
is not reasonable for the semantics of \surs which measures unshared road segments. 

For \erp and \neterp, we evaluate the query processing time with varying $\eta$ values. In order to make $\eta$ a 
dimensionless quantity (hence regardless of the unit of measurement, e.g., meter, km, mile...), for \erp, we scale down 
$\eta$ by $median(d(v, nn(v)))$, where $nn(v)$ is the nearest node of $v$ and $d$ is Euclidean distance, for \neterp, we 
scale down $\eta$ by $median(w(e))$, where $w(e)$ is the road length of edge $e$. We use median rather than minimum or 
maximum to mitigate the effect of outliers. 
Figure~\ref{fig:varying_eta} shows the results under several $(\tau_{\text{ratio}}, |Q|)$ settings on \beijing and 
\porto datasets. For \erp, we observe that small $\eta$ settings yield best overall performance. Although a 
larger $\eta$ is slightly faster in a few cases (e.g., $\eta = median(d(v,nn(v)))$ when $\tau_{\text{ratio}} = 0.3$ 
and $\size{Q} = 60$), the processing time rapidly increases when $\eta$ is larger than this choice. Hence we 
choose a small $\eta$ ($10^{-4} \cdot median(d(v,nn(v)))$) for consistently fast query processing. For \neterp, we 
choose $\eta = median(w(e))$ which yields the best performance for all the settings. 

\begin{figure}[t]
  \centering
  \includegraphics[width=0.9\linewidth]{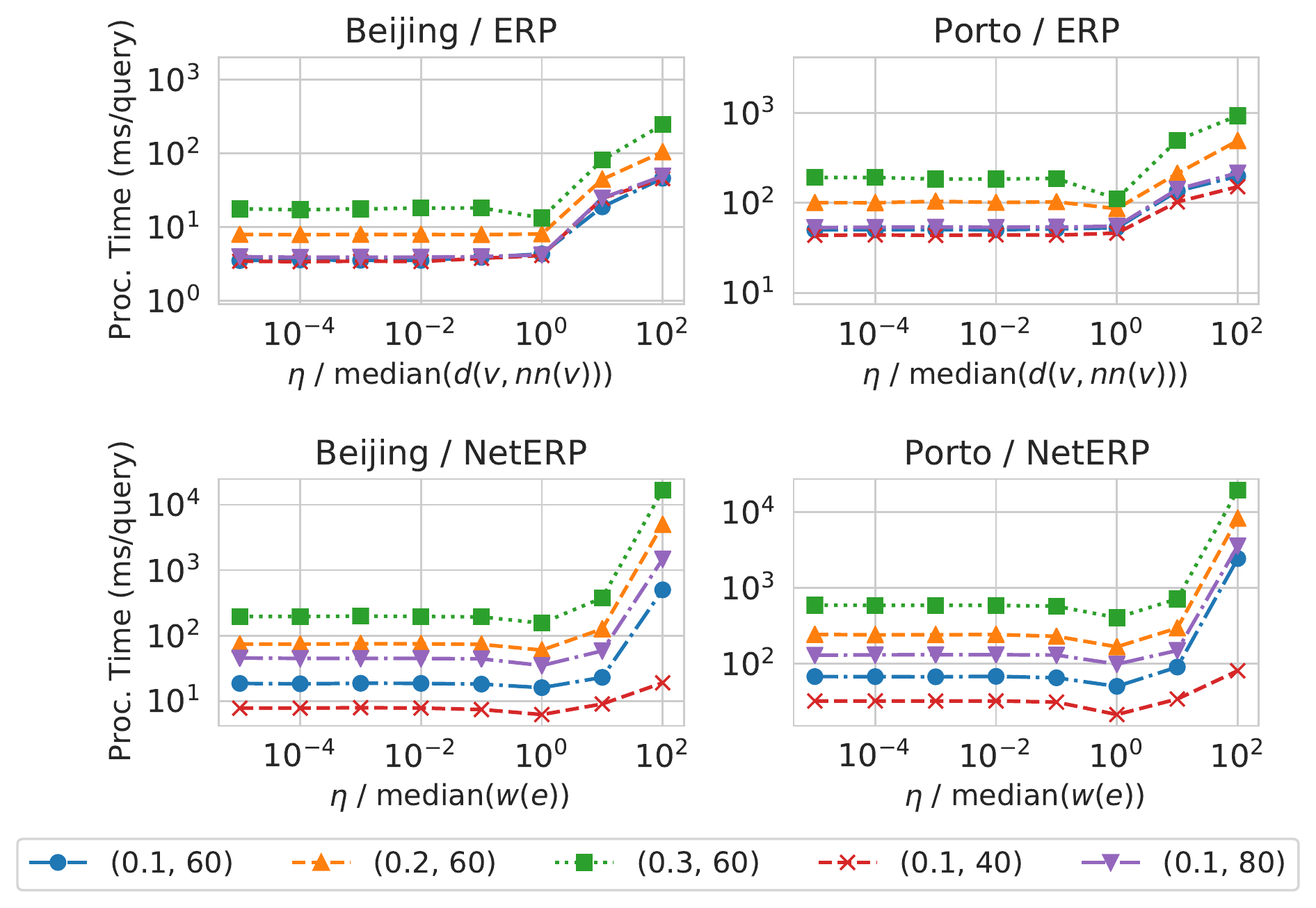}
  \caption{Varying $\eta$. Legend means $(\tau_\text{ratio}, |Q|)$.}
  \label{fig:varying_eta}
\end{figure}

\section{Detailed Setup for Effectiveness Evaluation}
\label{appendix:effectivenss}
We describe the detailed setup for travel time estimation in \S\ref{sec:experiments:travel-time}.
To estimate the travel time of a given path $Q$, we find a set of similar subtrajectories to $Q$ 
in the database $\database=\{(P^{(id)},T^{(id)})\}$. Here we call this set \emph{training dataset}.
Given a matched similar subtrajectory $P_{i:j}^{(id)}$, we can define the corresponding travel time 
by $T_j^{(id)}\!\!-T_i^{(id)}$. 
We estimate the travel time by averaging the travel time of the similar subtrajectories.
Note that, for a given $P^{(id)}$, there can be multiple subtrajectories that are similar to $Q$.
In case of that, we chose the most similar subtrajectory. In case of ties, we chose the shortest 
subtrajectory (otherwise \lors and \lcss become significantly worse because they 
do not penalize the length of trajectories). 

One issue here is how we evaluate the accuracy of the estimation, i.e., how to define the ground truth 
travel time. One option is using the average travel time of subtrajectories that exactly match to $Q$; 
however, this is problematic because travel time data used in this ground truth is also included in the 
training dataset. To avoid such data leakage, we employ cross-validation. Because the set of similar 
subtrajectories to $Q$ definitely includes the subtrajectories that exactly match $Q$, for each of these 
exact match subtrajectories $P^{(id)}_{i:j}$, we first exclude the travel time of $P^{(id)}_{i:j}$ from 
the training dataset, and then evaluate the accuracy of the average estimated travel time against the 
ground truth travel time. We repeat this process over all the subtrajectories that exactly match $Q$ and 
averaging the MSEs. The evaluation result is free from the data leakage discussed above.

To describe the above method formally, we first find the subtrajectories that exactly match $Q$. 
Suppose we have $N_Q$ exact match subtrajectories in the dataset. 
We compute the travel time for each of these exact match subtrajectoires, i.e., given a 
$P^{(id)}_{i:j} = Q$, we have $\omega:=T^{(id)}_j-T^{(id)}_i$. 
This is our ground truth. We denote the set of ground truths by 
$\Omega^{\text{exact}}:=\set{\omega_k\mid k=1 \ldots N_Q}$.

By leave-one-out cross-validation, the MSE of exact match is defined by 
\begin{align*}
  MSE(\text{exact}):=\frac{1}{N_Q}\sum_{k=1}^{N_Q}(\omega_k - avg(\Omega_{-k}^{\text{exact}}))^2, 
\end{align*}
where $\Omega_{-k}^{\text{exact}}$ is the set of ground truth data excluding the $k$-th travel time, 
i.e., $\Omega_{-k}^{\text{exact}}:=\Omega^{\text{exact}} \backslash \set{\omega_k}$. 

Then, we conduct similarity search with a threshold $\tau_{\ratio}$. We define 
$\Omega^{\tau_{\text{ratio}}}:=\set{T^{(id)}_j-T^{(id)}_i\mid\exists i,j \quad \text{s.t.\,} \wed(Q,P_{i:j}^{(id)})\le\tau_{\text{ratio}}\sum_{q \in Q}c(q)}$. 
For each data trajectory $P^{(id)}$, we choose only one subtrajectory $P_{i:j}^{(id)}$ that best matches $Q$, 
i.e., the one with the smallest WED to $Q$ (we break tie by picking the one with the shortest length). 
By leave-one-out cross-validation, the MSE of similarity search is defined by 
\begin{align*}
  MSE(\tau):=\frac{1}{N_Q}\sum_{k=1}^{N_Q}(\omega_k - avg(\Omega_{-k}^{\tau_{\text{ratio}}}))^2, 
\end{align*}
where $\Omega_{-k}^{\tau_{\text{ratio}}}$ is the set of estimated travel times excluding the $k$-the ground truth 
$\omega_k \in \Omega^{\text{exact}}$ (note that $\Omega^{\text{exact}} \subseteq \Omega^{\tau_{\text{ratio}}}$). 
Then RMSE is defined as $RMSE:=MSE(\tau) / MSE(\text{exact})$. 

We repeat the above procedure for each query $Q$ and report the average RMSE.

\section{Discussion on LORS, LCRS, and SURS}
\label{appendix:lors-lcrs-surs}
We show the following fact: if two trajectories $x$ and $y$ are similar in terms of \surs, then they are also similar in terms of \tsf{LCRS}~\cite{ICDE18-DISON}.
\tsf{LORS}~\cite{SIGIR18-Torch} and \tsf{LCRS} have the following relation
$$\tsf{LCRS}(x,y)=\frac{\tsf{LORS}(x,y)}{w(x)+w(y)-\tsf{LORS}(x,y)}$$
In addition, \tsf{LORS} and \surs have the following relation
$$\surs(x,y)=w(x)+w(y)-2\cdot\tsf{LORS}(x,y).$$

Suppose that $x$ and $y$ are similar in terms of \tsf{SURS}.
In this case, we have $w(x)+w(y)\simeq 2\cdot\tsf{LORS}(x,y)$. In addition, we also have $w(x)\simeq w(y)$ (because of the length filter). Therefore, we have $w(y)\simeq \tsf{LORS}(x,y)$ if $x$ and $y$ are similar in terms of \tsf{SURS}.
This leads to the following approximation:
$$1-\tsf{LCRS}(x,y)\simeq\frac{\tsf{SURS}(x,y)}{w(x)}$$
If $\tsf{SURS}(x,y)\le \tau_{ratio}\cdot w(x)$, then we approximately have $\tsf{LCRS}(x,y)\gtrapprox 1-\tau_{ratio}$. This is why \tsf{SURS} and \tsf{LCRS} behave similarly at small $\tau_{ratio}$. }

\end{document}